\documentclass[12pt]{article}
\usepackage{amsmath,amssymb,array,calc,rotating,epsfig,psfrag,amscd, cite}
\usepackage{epsfig}
\usepackage{amsmath}
\usepackage{braket}
\usepackage{tikz}
\usetikzlibrary{arrows}
\usepackage[utf8]{inputenc}

\makeatletter
\renewcommand\section{\@startsection {section}{1}{\z@}%
                                   {-3.5ex \@plus -1ex \@minus -.2ex}%nn
                                   {2.3ex \@plus.2ex}%
                                   {\normalfont\large\bfseries}}
\renewcommand\subsection{\@startsection{subsection}{2}{\z@}%
                                     {-3.25ex\@plus -1ex \@minus -.2ex}%
                                     {1.5ex \@plus .2ex}%
                                     {\normalfont\bfseries}}
\makeatother

\newcommand{\be}{\begin{equation}}
\newcommand{\ee}{\end{equation}}
\newcommand{\bea}{\begin{eqnarray}}
\newcommand{\eea}{\end{eqnarray}}

\newcommand{\al}{\alpha}

\newcommand{\s}{\sigma}

\newcommand{\C}{\mathbb{C}}

\newcommand{\hlf}{\frac{1}{2}}

\newcommand{\non}{\nonumber}

\newcommand{\Z}{\mathbb{Z}}
\newcommand{\D}{\mathcal{D}}
\newcommand{\Or}{\mathcal{O}}

\renewcommand{\O}{\operatorname{O}}
\newcommand{\SL}{\operatorname{SL}}

\newcommand{\Sp}{\operatorname{Sp}}

\newcommand{\Tr}{\operatorname{Tr}}

\newcommand{\lp}{\left(}
\newcommand{\rp}{\right)}
\newcommand{\ls}{\left[}
\newcommand{\rs}{\right]}

\newcommand{\ov}[1]{{\overline{#1}}}

\newcommand{\btau}{\bar{\tau}}

\tikzset{
	partial ellipse/.style args={#1:#2:#3}{
		insert path={+ (#1:#3) arc (#1:#2:#3)}
	}
}

\begin{document}
\begin{titlepage}

\begin{center}

%\today
\hfill November 14, 2019
%\hfill         MIFPA-14-03\phantom{xxx}

\vskip 2 cm
{\Large \bf Modular Orbits at Higher Genus}\\
\vskip 1.25 cm {Daniel Robbins\footnote{email address: dgrobbins@albany.edu} and Thomas Vandermeulen\footnote{email address: tvandermeulen@albany.edu}}\\

{\vskip 0.5cm \it Department of Physics, University at Albany, \\ Albany, NY 12222, USA \\}

\end{center}
\vskip 2 cm

\begin{abstract}
\baselineskip=18pt
We extend the modular orbits method of constructing a two-dimensional orbifold conformal field theory to higher genus Riemann surfaces.  We find that partition functions on surfaces of arbitrary genus can be constructed by a straightforward generalization of the rules that one would apply to the torus.  We demonstrate how one can use these higher genus objects to compute correlation functions and OPE coefficients in the underlying theory.  In the case of orbifolds of free bosonic theories by subgroups of continuous symmetries, we can give the explicit results of our procedure for symmetric and asymmetric orbifolds by cyclic groups.

\end{abstract}

\end{titlepage}

\pagestyle{plain}
%\baselineskip=18pt
% Try a wider skip
\baselineskip=19pt
%%%%%%%%%%%%%%%%%%%%%%%%%%%%%%%%%%%%%%%%%%%%%%%%%%%%%%%%%%%%%%%%%%%%%%%%%%%%%%
\section{Introduction}

Since the early days of two-dimensional Conformal Field Theory (CFT), it has been realized that the discipline has deep ties to the geometry of Riemann surfaces \cite{friedanshenker}.  In particular, bundles over the moduli spaces of these surfaces encode field theoretic data such as the spectrum and correlation functions.  By understanding the behavior of these objects in limiting situations (i.e.\ approaching boundaries of compactified moduli space), one can reconstruct field-theoretic quantities of interest.  In order to implement this in practice, however, we need some understanding of moduli spaces at arbitrary genus, their symmetries, and functions defined on them.  We also need specific field theories to investigate.

This brings us to the notion of an orbifold, which in CFT can be regarded as a mechanism by which we `divide out' symmetries to obtain a new theory from an old one \cite{Dixon:1985jw,Dixon:1986jc}.  In \cite{robbins2019orbifolds}, we proposed a procedure for constructing torus partition functions of orbifold theories which emphasizes modular invariance in favor of construction of twisted sector Hilbert spaces.  Such a method is firmly on the geometry side of this `field theory/geometry correspondence,' but by applying it to higher genus Riemann surfaces we will be able to recover field theoretic objects such as correlation functions and Operator Product Expansion (OPE) coefficients.  This paper describes the extension of the modular orbits method beyond genus one.

This is far from the first time that orbifolds have been considered on higher genus surfaces.  An excellent early exposition exists in \cite{DVV}, which focuses mainly on $\Z_2$ orbifolds of theories with central charge $c=1$.  Work such as~\cite{Tuite:2010yd,Tuite:2013bta} in the mathematical literature compute genus two partial traces for free fermion theories.  The technology employed in these works are appropriately specialized to their cases -- we aim to provide results that cover a broader class of theories.  

One motivation for extending the available higher genus technology comes from the modular bootstrap program (see~\cite{Cardy:1986ie,Cardy:1991kr,Hellerman:2009bu,Friedan:2013cba,Collier:2016cls,Afkhami-Jeddi:2019zci} and references therein).  Given a CFT, one can take its local data (spectrum and OPE coefficients), and construct the partition function on any given Riemann surface.  These partition functions must be invariant under modular transformations (the familiar $\SL(2,\Z)$ in the genus one case, $\Sp(2g,\Z)$ at genus $g$).  Turning this around, the requirement of modular invariance puts constraints on the local data of the CFT~\cite{Cardy:2017qhl,Keller:2017iql,Cho:2017fzo}.  By improving the tools needed to analyze CFTs on higher genus surfaces, and by increasing the list of examples which are manifestly modular invariant, as this paper aims to do, we hope to contribute to this ongoing research effort.

We begin in section \ref{review} by introducing the relevant material from the algebraic geometry of Riemann surfaces.  This includes a look at the partition functions of CFTs on these surfaces, notably the free bosonic theory which is solvable.  We also introduce the concept of a degenerating surface.  In section \ref{mainsection} we modify our proposed method of computing orbifold partition functions from \cite{robbins2019orbifolds} to include higher genus surfaces.  We follow this by using the Ising model as a toy example to demonstrate implementing the proposal and obtaining data about the underlying CFT from the higher genus partition function.  Section \ref{modtrans} examines the action of the mapping class and modular groups on higher genus surfaces, in particular their action on partial traces of the partition function.  With this knowledge, in section \ref{flavored} we compute higher genus partition functions for orbifolds of free bosonic theories by subgroups of continuous symmetries.  This allows us to implement our procedure in a fully explicit fashion; we demonstrate how one can compute correlation functions and OPE coefficients in the resulting theories.  Finally we conclude in section \ref{conclusion} with a summary and outlook at further applications of these methods.

\section{Review of Higher Genus Riemann Surfaces}
\label{review}

A standard reference for the material presented here is \cite{fay1973theta}.  On a Riemann surface of genus $g$ we have a homology basis consisting of $2g$ cycles, traditionally called $a_i$ and $b_i$ ($i=1,...,g$).  Correspondingly, the cohomology has a basis in $g$ holomorphic and $g$ antiholomorphic one-forms, $\omega_i(z)$ and $\bar{\omega}_i(\bar{z})$.  Conformal invariance allows us to fix their $a$ periods, while their $b$ periods form the moduli $\tau_{ij}$ that describe our surface.  We can summarize this as
\be
\oint_{a_i}\omega_j=\delta_{ij}\hspace{.5cm}\oint_{b_i}\omega_j=\tau_{ij}.
\ee
with conjugate relations for $\bar{\omega}_i(\bar{z})$.  Generalizing the complex structure constant $\tau$ from the torus, the period matrix $\tau_{ij}$ is symmetric with positive-definite imaginary part.  It gives a space of complex dimension $3g-3$.  For $g\le 3$ the entries of this matrix can be taken directly as the moduli describing the surface.  Above genus three the correspondence ceases to be one-to-one; the unconstrained entries in the period matrix exceeds the number of moduli, and we must impose constraints known as Schottky relations on $\tau_{ij}$.

We are able to define theta functions associated with these surfaces, as well.  The higher genus equivalent to the usual theta function is the Siegel theta function, given by\footnote{Here and going forward, we use a dot to denote the contraction of multi-component objects, e.g. $x\cdot\tau\cdot x$ should be understood as $\sum_{i,j=1}^gx_i\tau_{ij}x_j$.  We also will tend to omit indices, writing the genus $g$ period matrix simply as $\tau$.}
\be
\label{thetafunc}
\theta(z|\tau)=\sum_{x\in\Z^g}\exp{\left[\pi ix\cdot\tau\cdot x+2\pi ix\cdot z\right]}.
\ee
Here $z$ is a $g$-vector.  Luckily we have a canonical way of associating points $y$ on our Riemann surface with $g$-vectors; we use the Abel map, given by
\be
z_i(y)=\int_{y_0}^y\omega_i.
\ee
This construction maps our surface onto its Jacobian variety, the complex $g$-torus given by $\C^g/(\Z^g+\tau\Z^g)$.  This allows us to regard (\ref{thetafunc}) as a function of a point on our surface and its period matrix, in analogy to the situation on the torus.

Assigning antiperiodic boundary conditions to the cycles of our torus has the effect of shifting $z$ and $x$ by half lattice vectors.  Since this is a situation that arises often, we call these quantities theta functions with characteristics (also known as spin structures), and write them as
\be
\label{thetafuncchar}
\theta\left[\begin{matrix}\delta \\ \epsilon\end{matrix}\right](z|\tau)=\sum_{x\in\Z^g}\exp{\left[\pi i\left(x+\delta\right)\cdot\tau\cdot\left(x+\delta\right)+2\pi i\left(x+\delta\right)\cdot\left(z+\epsilon\right)\right]}.
\ee
The usual choice of characteristics is to take $(\delta,\epsilon)\in \frac{1}{2}\Z_2^g\times\frac{1}{2}\Z_2^g$ i.e. they are $g$-component vectors whose entries are each 0 or 1/2.  In this case, we see that the parity of a half-integer characteristic theta function in its $z$ argument is given by the quantity $4(\delta\cdot\epsilon)$ mod 2 (0 for even, 1 for odd).  By extension we label the characteristic as even or odd.  A theta function (with characteristics) evaluated at $z=0$ is known as a \textit{theta constant}; one sees from the preceding discussion of periodicity that odd theta constants vanish.  When we have an odd characteristic, which we will write as $\Delta=(\delta,\epsilon)$ for short, it is sensible to define a spinor as
\be
\label{spinor}
h_\Delta(z|\tau)=\left[\sum_{i=1}^g\partial_{z_i}\theta_\Delta(0|\tau)\omega_i(z)\right]^{1/2}.
\ee
Finally, we write the prime form,
\be
\label{primeform}
E(z,w)=\frac{\theta_\Delta(\int^z_w\omega|\tau)}{h_\Delta(z|\tau)h_\Delta(w|\tau)}
\ee
which (as the notation indicates) is independent of the particular choice of odd characteristic $\Delta$.

As an example, consider the torus.  There we have three even spin structures $(0,0)$, $(0,\frac{1}{2})$, $(\frac{1}{2},0)$ and a unique odd spin structure, $(\frac{1}{2},\frac{1}{2})$.  One often writes these four functions as
\be
\theta_1(z|\tau)\equiv\theta\genfrac[]{0pt}{0}{\frac{1}{2}}{\frac{1}{2}}(z|\tau),\hspace{.25cm}\theta_2(z|\tau)\equiv\theta\genfrac[]{0pt}{0}{\frac{1}{2}}{0}(z|\tau),\hspace{.25cm}\theta_3(z|\tau)\equiv\theta\genfrac[]{0pt}{0}{0}{0}(z|\tau),\hspace{.25cm}\theta_4(z|\tau)\equiv\theta\genfrac[]{0pt}{0}{0}{\frac{1}{2}}(z|\tau). 
\ee
The list of identities that these functions satisfy is extensive.  We mention one that will appear repeatedly in our calculations:
\be
\label{thetaetaidentity}
\partial_z\theta_1(0|\tau)=-2\pi\eta^3(\tau).
\ee

We can choose the coefficient of $dz$ for the single holomorphic one-form $\omega$ to be a constant (which we take as 1), so at genus one the spinor (\ref{spinor}) is constant over the surface and takes the form
\be
h_\Delta=\sqrt{\theta'_1(0|\tau)}.
\ee
The torus prime form is then
\be
E(z,w)=\frac{\theta_1(z-w|\tau)}{\theta'_1(0|\tau)}.
\ee

\subsection{Partition Functions}

What should we expect from the partition function of a CFT evaluated on such a surface?  We illustrate the case of a single scalar field to gain intuition and prepare for more detailed examples later.  Recall that the partition function, or vacuum amplitude, of a theory is simply the path integral evaluated with no additional operator insertions:
\be
\label{Z}
Z=\int\D\varphi e^{-S_E[\varphi]},
\ee
with $S_E[\varphi]$ the Euclidean (Wick rotated) action functional.  Recall that, in order to compute the torus partition function, we would have imposed the following boundary conditions on our field
\be
\label{torusbcs}
\varphi(z+1,\bar{z}+1)=\varphi(z,\bar{z}),\hspace{.5cm}\varphi(z+\tau,\bar{z}+\bar{\tau})=\varphi(z,\bar{z}),
\ee
i.e. it should obey the periodicities of the surface it's defined on.  On a higher genus surface we impose similar conditions
\be
\label{highergbcs}
\varphi(z+a_i,\bar{z}+a_i)=\varphi(z,\bar{z}),\hspace{.5cm}\varphi(z+b_i,\bar{z}+b_i)=\varphi(z,\bar{z}).
\ee
This expression should be understood schematically as imposing periodicity on the fields as their arguments wind the various $a$ and $b$ cycles of the higher genus surface.

In the case of the free scalar, we are used to identifying one cycle as space and the other as time.  The winding numbers along these cycles give the familiar momentum and winding of the boson.  Topologically, the path integral (\ref{Z}) will have instanton contributions from paths that wind these cycles.  We refer to this as the momentum lattice part of the partition function, and it is given by
\be
Z_{\text{mom.}}=(\text{det Im}\tau)^{1/2}\sum_{(p_L,p_R)\in\Gamma_g}\exp{\left[\frac{2\pi i}{4}\left(p_L\cdot\tau\cdot p_L-p_R\cdot\bar{\tau}\cdotp_R\right)\right]}
\ee
where the momenta live on the lattice
\be
\Gamma_g=\left\{\left(\frac{x}{R}+yR,\frac{x}{R}-yR\right)\bigg|(x,y)\in\Z^g\times\Z^g\right\},
\ee
with $R$ the radius of compactification.  The remaining part of the path integral handles the contributions of oscillator modes, so we refer to it as the oscillator piece.  Having no way to detect the winding, this piece is identical (up to an infinite multiplicative constant from the noncompact zero mode) to the partition function of the noncompact boson.  The integral can be evaluated by noting that the boson action is gaussian, and the result is
\be
\label{zosc}
Z_{\text{osc.}}=(\text{det Im}\tau)^{-{1/2}}(\text{det}\Delta)^{-1/2}
\ee
where $\Delta$ is the scalar Laplacian, whose determinant is understood to be zeta function regularized.  Combining these two gives the full partition function at genus $g$ as \cite{evthesis}
\be
\label{genusgz}
Z=\frac{1}{\sqrt{\det{\Delta}}}\sum_{(p_L,p_R)\in\Gamma_g}\exp{\left[\frac{2\pi i}{4}\left(p_L\cdot\tau\cdot p_L-p_R\cdot\bar{\tau}\cdot p_R\right)\right]}.
\ee

Some comments on the prefactor $(\det{\Delta})^{-1/2}$ are in order.  Na\"{i}vely, one would expect that the determinant of the Laplacian would factorize into the product of determinants of chiral Dirac operators.  However, the conformal anomaly gives an obstruction to this factorization in the form of the Liouville action $S_L$.  We expect a relation of the form \cite{VV}
\be
\det{\Delta}=e^{cS_L}|\det{\partial_0}|^2,
\ee
where $\partial_0$ is the chiral Dirac operator acting on scalars and $c$ the theory's central charge.  The quantity $\det{\partial_0}$ is expected to be well-behaved under the degeneration relations defined in the following section, in the sense that, in the leading order, it simply goes to its lower genus counterpart(s) \cite{VV}.  The anomalous term $e^{cS_L}$, however, is dependent on the metric chosen for the higher genus surface (it does not show up on the torus because we can always choose a flat metric).  Its precise form will have no bearing on the CFT data we wish to obtain.

One way to effectively disregard this term is to take the quotient of the partition function in question with the appropriate power of the noncompact free boson partition function (the power being such that the central charges match) \cite{Behera:1989gg}.  Since the anomalous terms are universal, they will cancel.  We then apply degeneration to the result and, knowing the results for the boson, extract the information we desire about the CFT of interest.  To this end, going forward we will write the prefactor $(\det{\Delta})^{-1/2}$ as a function $H_g(\tau)$ on the moduli space.  The only details of $H$ we will need are that it goes to its lower genus counterpart(s) under degeneration and that $H_1(\tau)=|\eta(\tau)|^{-2}$.

\subsection{Degeneration}
\label{degensec}

An important operation on higher genus Riemann surfaces is degeneration, in which a surface tends toward a point on the boundary of its moduli space where it approaches a surface (or surfaces) of lower genus.  More precisely, for any surface $g\ge 2$ we have the \textit{separating degeneration} in which a cycle that is trivial in homology tends to zero.  In this limit, our surface of genus $g$ resembles two surfaces of genera $g_1+g_2=g$ connected by a long, thin tube.  Degeneration provides a link between data on surfaces of different genera, and allows us to extract CFT data beyond the spectrum from higher genus partition functions.  

Now we specialize to the case of a surface of genus two degenerating to two tori, with complex structure constants $\tau_1$ and $\tau_2$.  In particular, when we parameterize the separating degeneration by a parameter $t\in\C\to 0$,
the period matrix has a $t$ expansion of the form \cite{MT}
%\begin{multline}
\be
\label{pmsepdegen}
\tau\to\left(\begin{matrix}\tau_1 & 0 \\ 0 & \tau_2\end{matrix}\right)+2\pi it\left(\begin{matrix}0 & 1 \\ 1 & 0\end{matrix}\right)+\O(t^2).
\ee
%+(2\pi it)^2\left(\begin{matrix}-2\partial\log{\eta(\tau_2)} & 0 \\ 0 & -2\partial\log{\eta(\tau_1)}\end{matrix}\right) \\
%+(2\pi it)^3\left(\begin{matrix}0 & -4\partial\log{\eta(\tau_1)}\partial\log{\eta(\tau_2)} \\ -4\partial\log{\eta(\tau_1)}\partial\log{\eta(\tau_2)} & 0\end{matrix}\right)+\O(t^4).
%\end{multline}
%We have written the above expression up to order $t^3$ because, as we will see shortly, this will be the minimum required order to obtain nontrivial correlation functions from this degeneration.  
In general, by inserting a complete set of states in the long, thin tube, we would expect the partition function to have an expansion in $t$ of the form~\cite{evthesis}
\be
\label{separatingschematic}
%Z\to\sum_{h_i,\bar{h}_j}t^{h_i}\bar{t}^{\bar{h_j}}\braket{\Or_i(0)}^{\tau_1}\braket{\Or_j(0)}^{\tau_2},
Z\to\sum_{\mathrm{operators\ }i}t^{h_i}\bar{t}^{\bar{h_i}}\braket{\Or_i(0)}^{\tau_1}\braket{\Or_i(0)}^{\tau_2},
\ee
where the sum is over weights of operators appearing in the theory.  The superscript of the correlation function denotes the complex structure constant of the surface on which it has been evaluated.  As a check, the vacuum is the lowest weight state with $h=\bar{h}=0$, and so at lowest order the genus two partition function indeed separates into the product of the two genus one partition functions.

The other type of degeneration is one in which we let a homologically nontrivial cycle degenerate.  This causes our surface of genus $g$ to approach a surface of genus $g-1$ with a thin handle attached.  Again specializing to genus two, we will use translation invariance to set the location of one end of the thin handle to 0; the location of the other will be called $z$.  Again we parameterize the degeneration in terms of a complex parameter $t$.  The form of the period matrix under this degeneration is
\be
\label{pmnonsepdegen}
\tau\to\left(\begin{matrix}\tau & z \\ z & \frac{1}{2\pi i}\log{\left[\frac{t}{E^2(z,0)}\right]}\end{matrix}\right)+
\O(t),
%\frac{t}{\pi i}\left(\begin{matrix}1 & -\partial_z\log{\theta_1(z|\tau)} \\ -\partial_z\log{\theta_1(z|\tau)} & (\partial_z\log{\theta_1(z|\tau)})^2\end{matrix}\right)+O(t^2)
\ee
where $E(z,0)$ is the prime form (\ref{primeform}).  As before, we have an expectation for the form of the partition function's behavior~\cite{evthesis}:
\be
\label{nonseparatingschematic}
%\sum_{h_i,\bar{h}_i}t^{h_i}\bar{t}^{\bar{h_i}}\braket{\Or_i(z)\Or_i(0)},
\sum_{\mathrm{operators\ }i}t^{h_i}\bar{t}^{\bar{h_i}}\braket{\Or_i(z)\Or_i(0)}^\tau,
\ee
where now we find that we are calculating genus-one two-point functions of operators with themselves.

\section{Orbifolds at Higher Genus}
\label{mainsection}

In an orbifold theory, we define partial traces similarly to the partition function (\ref{Z})
\be
\label{partialtrace}
Z_{k,g}(\tau,\btau)=\int\mathcal{D}\varphi_{k,g}e^{-S_E[\varphi]},
\ee
except we have modified our boundary conditions from (\ref{torusbcs}) to include transformation by group elements in the periodicity:
\be
\label{twistedtorusbcs}
\varphi(z+1,\bar{z}+1)=k\cdot\varphi(z,\bar{z}),\hspace{.5cm}\varphi(z+\tau,\bar{z}+\bar{\tau})=g\cdot\varphi(z,\bar{z}).
\ee

At higher genus we would make an analogous change to (\ref{highergbcs}), which would give us the boundary conditions
\be
\varphi(z+a_i,\bar{z}+a_i)=k_i\cdot\varphi(z,\bar{z}),\hspace{.5cm}\varphi(z+b_i,\bar{z}+b_i)=g_i\cdot\varphi(z,\bar{z}),
\ee
which we again emphasize are to be understood schematically.  The partial traces that we compute would now be of the form
\be
\label{pthigher}
Z_{k_1,...,k_g;g_1,...,g_g}(\tau,\btau)=\int\mathcal{D}\varphi_{k_1,...,k_g;g_1,...,g_g}e^{-S_E[\varphi]},
\ee
where we are imposing group element boundary conditions on each of the surface's $2g$ homotopy one-cycles.

We begin with a proposal to construct partition functions of $\Z$ or $\Z_N$ orbifold theories.  In this situation, since the orbifold groups are abelian, we do not have to worry about imposing commutation constraints on the elements appearing in the partial traces (\ref{pthigher}).  Also, since cyclic groups have $H^2(G,U(1))=0$, we should not have disconnected orbits entering with a choice of discrete torsion.  Nevertheless, we discuss a simple case where discrete torsion does arise, namely $\Z_2\times\Z_2$, in the context of orbit structure at higher genus in section \ref{z2z2}.  The steps given in \cite{robbins2019orbifolds} for the analogous situation on the torus generalize in a straightforward way to higher genus surfaces:
\begin{enumerate}
	\item Use the knowledge of the parent theory to construct the untwisted sector partial traces $Z_{0,...,0;n_1,...,n_g}$.
	\item Apply modular transformations to the untwisted sector partial traces to obtain all partial traces $Z_{m_1,...,m_g;n_1,...,n_g}$.  Note that (for the $\Z_N$ case) the subscripts may not be periodic modulo $N$ (but will be periodic modulo $N^2$).
	\item Construct the twisted sector partition functions
	\be
	Z_{m_1,...,m_g}(\tau,\btau)=\frac{1}{N^{2g}}\sum_{n_1=0}^{N^2-1}...\sum_{n_g=0}^{N^2-1}Z_{m_1,...,m_g;n_1,...,n_g}(\tau,\btau).
	\ee
	Here the $m_i$ will be periodic modulo $KN$ for some integer $1\le K\le N$, and we can construct the full orbifold partition function as
	\be
	Z_G(\tau,\btau)=\sum_{m_1=0}^{KN-1}...\sum_{m_g=0}^{KN-1}Z_{m_1,...,m_g}(\tau,\btau).
	\ee
\end{enumerate}

Before diving into specific examples, we can extract new features from the material presented so far.  Consider how partial traces behave in the degeneration limits of section \ref{degensec}.  For simplicity consider a genus two partition function under the separating degeneration.  We expect it to yield a series of the form (\ref{separatingschematic}).  We could write an analogous expression for a partial trace $Z_{k_1,k_2;g_1,g_2}$:
\be
\label{ptsepschematic}
%Z_{k_1,k_2;g_1,g_2}\to\sum_{h_i,\bar{h}_j}t^{h_i}\bar{t}^{\bar{h_j}}\braket{\Or_i(0)}^{\tau_1}_{k_1,g_1}\braket{\Or_j(0)}^{\tau_2}_{k_2,g_2}.
Z_{k_1,k_2;g_1,g_2}\to\sum_{\mathrm{operators\ }i}t^{h_i}\bar{t}^{\bar{h_i}}\braket{\Or_i(0)}^{\tau_1}_{k_1,g_1}\braket{\Or_i(0)}^{\tau_2}_{k_2,g_2}.
\ee
What are these $\braket{\Or(0)}_{k,g}$ that have appeared?  When $h=\bar{h}=0$ they are the partial traces of the partition function on the torus, but in higher orders they are objects we have not yet examined.  Following the rest of the proposed orbifold procedure, we would calculate the full genus two partition function by taking a sum over modular orbits of the expression (\ref{ptsepschematic}).  Finally, expanding the left hand side as a series in $t$, we see from interchanging the modular orbit and degeneration operations that the one-point function of the operator $\Or$ in the orbifold theory is given (just as with the vacuum one-point function) by a sum over modular orbits of the object $\braket{\Or(0)}_{k,g}$.  This suggests that we should identify
\be
\label{ptor}
\braket{\Or(z)}_{k,g}=\int\D\varphi_{k,g}\Or(z)e^{-S_E[\varphi]}
\ee
where the subscript on the measure indicates the group-twisted boundary conditions (\ref{twistedtorusbcs}).  This expression is written for the one-point function of an operator on the torus, but is straightforwardly extended to encompass multi-point functions of various operators at arbitrary genus.  One would use (\ref{ptor}) in place of the partition function partial traces $Z_{k,g}$ to calculate correlation functions of operators that survive the orbifold procedure (that is, operators which were present in both the parent theory and the resulting orbifold theory).

\subsection{Example: Ising Model Partition Function}
\label{isingorbifold}

As a simple example of the principles we've laid out so far, consider the Ising model.  Its partition function can be written at any genus as \cite{DVV}
\be
\label{zising}
Z_{\text{Ising}}^{(g)}(\tau)=H_g^{1/2}(\tau)2^{-g}\sum_{\alpha,\beta}\left|\theta\genfrac[]{0pt}{0}{\alpha}{\beta}(0|\tau)\right|
\ee
where the sum is over all half-integer characteristics.  Viewed as a minimal model, it has three primary states, the vacuum $|1\rangle$ with $h=\bar{h}=0$, a state $|\epsilon\rangle$ with $h=\bar{h}=1/2$, and a state $|\s\rangle$ with $h=\bar{h}=1/16$.  
%States in generic minimal models can be characterized by three integers: $p,n,m$.  
%The weights $h^{(p)}_{n,m}=\bar{h}^{(p)}_{n,m}$ of the three Ising model states are given by $h_1=0$, $h_{2,1}^{(3)}=1/2$ and $h_{2,2}^{(3)}=1/16$.  
The model possesses a symmetry under which $|\s\rangle$ changes sign and the other two states remain invariant.  In order to construct the partition function of an orbifold by this symmetry, it will help to rewrite (\ref{zising}) on the torus in terms of minimal model characters:
\be
\label{isingmmz}
Z_{\text{Ising}}^{g=1}=\left|\chi_1\right|^2+\left|\chi_\epsilon\right|^2+\left|\chi_\s\right|^2,
\ee
where the $\chi_i(\tau)$ are the characters for the $p=3$ minimal model.  
%\be
%\chi^{(p)}_{n,m}=\frac{1}{\eta(q)}\ls\sum_{k\in\Z}q^{p(p+1)\lp k+\frac{(p+1)n-pm}{2p(p+1)}\rp^2}-\sum_{k\in\Z}q^{p(p+1)\lp k+\frac{(p+1)n+pm}{2p(p+1)}\rp^2}\rs.
%\ee
We will need the transformation rules of these characters under modular transformations,
\be
\chi_1(\tau+1)=\zeta\chi_1(\tau),\quad\chi_\epsilon(\tau+1)=-\zeta\chi_\epsilon(\tau),\quad\chi_\s(\tau+1)=e^{\pi i/8}\zeta\chi_\s(\tau),
\ee
where $\zeta=e^{-\pi i/24}$, and
\be
\lp\begin{matrix}\chi_1(-1/\tau) \\ \chi_\epsilon(-1/\tau) \\ \chi_\s(-1/\tau)\end{matrix}\rp=\lp\begin{matrix} \hlf & \hlf & \frac{1}{\sqrt{2}} \\ \hlf & \hlf & -\frac{1}{\sqrt{2}} \\ \frac{1}{\sqrt{2}} & -\frac{1}{\sqrt{2}} & 0 \end{matrix}\rp\lp\begin{matrix} \chi_1(\tau) \\ \chi_\epsilon(\tau) \\ \chi_\s(\tau)\end{matrix}\rp.
\ee

One can check that on the torus (\ref{zising}) and (\ref{isingmmz}) are in fact equivalent.  As a review of our procedure at genus one, we construct the partition function of the Ising model orbifold by its $\Z_2$ symmetry.

As we understand the symmetry by its effect on the primary states, we can quickly write the untwisted sector partial traces (using multiplicative notation for $\Z_2$) as
\bea
Z_{1,1}(\tau,\btau) &=& \left|\chi_1\right|^2+\left|\chi_\epsilon\right|^2+\left|\chi_\s\right|^2,\\
Z_{1,-1}(\tau,\btau) &=& \left|\chi_1\right|^2+\left|\chi_\epsilon\right|^2-\left|\chi_\s\right|^2.
\eea
Applying the method of modular orbits we can generate
\bea
Z_{-1,1}(\tau,\btau) &=& Z_{1,-1}(-1/\tau,-1/\btau)\non\\
&=& \left|\hlf\chi_1+\hlf\chi_\epsilon+\frac{1}{\sqrt{2}}\chi_\s\right|^2+\left|\hlf\chi_1+\hlf\chi_\epsilon-\frac{1}{\sqrt{2}}\chi_\s\right|^2 -\left|\frac{1}{\sqrt{2}}\chi_1-\frac{1}{\sqrt{2}}\chi_\epsilon\right|^2\non\\
&=& \chi_1\ov{\chi}_\epsilon+\chi_\epsilon\ov{\chi}_1+\left|\chi_\s\right|^2,\\
Z_{-1,-1}(\tau,\btau) &=& Z_{-1,1}(\tau+1,\btau+1)\non\\
&=& -\chi_1\ov{\chi}_\epsilon-\chi_\epsilon\ov{\chi}_1+\left|\chi_\s\right|^2.
\eea
At this point we're done, and we can examine the resulting orbifold partition function.  We can work by sector; the untwisted sector partition function is
\be
Z_1=\hlf\lp Z_{1,1}+Z_{1,-1}\rp=\left|\chi_1\right|^2+\left|\chi_\epsilon\right|^2,
\ee
which simply consists of the invariant states $|1\rangle$ and $|\epsilon\rangle$ and their descendants.  The twisted sector partition function is
\be
Z_{-1}=\hlf\lp Z_{-1,1}+Z_{-1,-1}\rp=\left|\chi_\s\right|^2,
\ee
so this consists simply of the state $|\s\rangle$ and its descendants.  The full partition function is just the original Ising model partition function back again.

In order to show that our proposed formalism yields the expected results at higher genus as well, we calculate by modular orbits the same result at genus two.  This computation is considerably lengthier than its torus counterpart, so we leave the full details to appendix \ref{isingorbg2}.  The result is, as expected, that we recover the Ising partition function, and the orbifold acts trivially.

\subsection{Example: Ising Model Correlation Functions}
\label{isingcorr}

Ising also provides a convenient check on the method of degeneration, since we know both its genus two partition function and genus one correlation functions.  We begin with the separating degeneration, using the form of $\tau$ given in (\ref{pmsepdegen}) to rewrite (\ref{zising}) as a series in $t$.  We see that, to obtain the lowest order term in this series, we can simply take $t\to 0$.  In that case, the genus two theta function reduces to a product of two genus one theta functions.  So, we have
\be
Z_{\text{Ising}}^{(2)}\to \frac{1}{4}H_1^{1/2}(\tau_1)H_1^{1/2}(\tau_2)\sum_{\substack{\alpha_1,\beta_1\in \Z_2^2 \\ \alpha_2,\beta_2\in \Z_2^2}}\left|\theta\genfrac[]{0pt}{0}{\alpha_1}{\beta_1}(0|\tau_1)\theta\genfrac[]{0pt}{0}{\alpha_2}{\beta_2}(0|\tau_2)\right|=Z_{\text{Ising}}^{(1)}(\tau_1)Z_{\text{Ising}}^{(1)}(\tau_2).
\ee
As expected, to lowest order we obtain the product of the correctly normalized torus partition functions.  Let us also calculate the next term, which will be useful for the calculation of OPE coefficients.  We get the coefficient of the order-$t$ term in the theta function's Taylor series as
\be
\partial_z\theta\genfrac[]{0pt}{0}{\alpha_1}{\beta_1}(0|\tau_1)\partial_z\theta\genfrac[]{0pt}{0}{\alpha_2}{\beta_2}(0|\tau_2),
\ee
which will be nonzero only when both genus one spin structures are odd i.e. $\alpha_1=\alpha_2=\beta_1=\beta_2=1/2$.  For this term in the sum over spin structures, the holomorphic and antiholomorphic leading parts cancel due to the vanishing of odd theta constants.  This allows us to obtain a $|t|$ term, given by
\be
|t|\pi^2|\eta(\tau_1)|^2|\eta(\tau_2)|^2.
\ee
Comparing this result to (\ref{separatingschematic}), we identify the above term as giving the torus one-point function of the $h=\bar{h}=1/2$ primary field $\varepsilon$:
\be
\label{epsilon1pt}
\braket{\varepsilon(0)}=\pi |\eta(\tau)|^2.
\ee
This result agrees with other methods of calculation \cite{DiFrancesco:1987ez}.

We now move to the non-separating degeneration limit of our genus two surface.  Inserting (\ref{pmnonsepdegen}) into (\ref{zising}), we again obtain a series in $t$:
\be
\frac{1}{4}|\eta(\tau)|^{-1}\sum_{\substack{\alpha_1,\beta_1\in \Z_2^2 \\ \alpha_2,\beta_2\in \Z_2^2}}\left|\sum_{x_2\in\Z}\left[\frac{t}{E^2(z,0)}\right]^{\frac{(x_2+\alpha_2)^2}{2}}\theta\genfrac[]{0pt}{0}{\alpha_1}{\beta_1}((x_2+\alpha_2)z|\tau)e^{2\pi i\beta_2(x_2+\alpha_2)}\right|.
\ee
The lowest order term has $x_2=\alpha_2=0$, in which the $\beta_2$ sum gives simply a factor of 2 and we recover the partition function.  

At the next order we have contributions from the $\alpha_2=1/2$ terms both when $x_2=0$ and $x_2=-1$.  Writing out these terms along with the terms from the two values of $\beta_2$, the remaining sum over genus one spin structure takes the form
\be
\sum_{\alpha_1,\beta_1\in\Z_2^2}\left|\theta\genfrac[]{0pt}{0}{\alpha_1}{\beta_1}\left(\frac{z}{2}|\tau\right)+\theta\genfrac[]{0pt}{0}{\alpha_1}{\beta_1}\left(-\frac{z}{2}|\tau\right)\right|+\left|\theta\genfrac[]{0pt}{0}{\alpha_1}{\beta_1}\left(\frac{z}{2}|\tau\right)-\theta\genfrac[]{0pt}{0}{\alpha_1}{\beta_1}\left(-\frac{z}{2}|\tau\right)\right|.
\ee
Whether the spin structure is odd or even these terms collapse to the same result, so our next term in the $t$-series can be written as
\be
\frac{1}{2}\left|\frac{t}{E^2(z,0)}\right|^{1/8}|\eta(\tau)|^{-1}\sum_{\alpha,\beta\in\Z_2^2}\left|\theta\genfrac[]{0pt}{0}{\alpha}{\beta}\left(\frac{z}{2}|\tau\right)\right|.
\ee
Comparing with (\ref{nonseparatingschematic}), we can identify the torus two-point function of the field $\sigma$ (which has $h=\bar{h}=1/16$) with itself, again agreeing with expectation \cite{DVV}:
\be
\label{sigmasigma2pt}
\braket{\sigma(z)\sigma(0)}=\frac{1}{2}|E(z,0)|^{-1/4}|\eta(\tau)|^{-1}\sum_{\alpha,\beta\in\Z_2^2}\left|\theta\genfrac[]{0pt}{0}{\alpha}{\beta}\left(\frac{z}{2}|\tau\right)\right|.
\ee

We note that a two-point correlation function of an operator with itself in CFT is expected, when expanded in $z$, to take the form
\be
\label{opeexpansion}
\braket{\Or_i(z)\Or_i(0)}\sim \sum_j\lambda_{iij}z^{h_j-2h_i}\bar{z}^{\bar{h}_j-2\bar{h}_i}\braket{\Or_j(0)}.
\ee
Setting $z=0$ in (\ref{sigmasigma2pt}) we indeed obtain the vacuum one-point function (partition function) with unit coefficient, so we have $\lambda_{\sigma\sigma 1}=1$.  Moving to the next order term, we expand the theta function in $z$.  Using the fact that $E(z,0)$ is linear in $z$ to leading order and evaluating the theta derivative in terms of eta functions, the next term will be
\be
\frac{1}{2}z^{1/2-1/8}\bar{z}^{1/2-1/8}\pi |\eta(\tau)|^2.
\ee
Comparing with (\ref{opeexpansion}), we learn two things.  First we see that $\lambda_{\sigma\sigma\sigma}=0$ since expanding our theta sum did not yield a $|z|^{1/8}$ term.  This is consistent with the expectation that the multi-point function of an odd number of $\sigma$ vanishes (note that we didn't find a $\sigma$ one-point function when we applied the separating degeneration).  Using our earlier result (\ref{epsilon1pt}) for $\braket{\varepsilon(0)}$, we also learn that $\lambda_{\sigma\sigma\varepsilon}=1/2$, which agrees with the literature \cite{YellowBook}.

\section{Modular Transformations}
\label{modtrans}

The large diffeomorphisms of a Riemann surface (i.e. diffeomorphisms modulo those smoothly connected to the identity) form a group known as the \textit{mapping class group}.  For the torus this is the familiar modular group SL(2;$\Z$), but at higher genera it has a more involved structure.  Its action on the period matrix, however, is easy to describe.  Our homology basis of $a$ and $b$ cycles has an (antisymmetric) intersection product $\circ$ given by
\be
\label{homologybasis}
a_i\circ b_j=\delta_{ij},\hspace{.5cm}a_i\circ a_j=b_i\circ b_j=0.
\ee
Modular transformations are given by linear transformations of the one-cycles which preserve (\ref{homologybasis}), which necessitates that they be given by the symplectic group Sp$(2g;\Z)$.  This action carries over to the period matrix, and indeed the partition functions we've examined so far are invariant under such transformations of $\tau$.  Orbifold partial traces, however, are sensitive to the full mapping class group, which is an extension of Sp$(2g;\Z)$ \cite{massuyeau}.  In this section we examine the action of this group on the homotopy one-cycles of a genus two surface, which will allow us to deduce its effect on partial traces.

\subsection{Torus Review}

Let's begin by understanding the torus' SL(2;$\Z$) symmetry in a language that will readily generalize to higher genus.  Homologically the torus has two independent cycles, which we'll call $a$ and $b$.  There is also a holomorphic one-form $\omega$.  We can choose to normalize $\omega$ such that
\be
\int_a\omega=1\hspace{1cm} \int_b\omega=\tau
\ee
with $\tau\in\mathbb{H}$ being the familiar complex structure constant.  The mapping class group is generated by Dehn twists about these cycles, which involves cutting the surface along a given cycle, rotating one side of the cut by $2\pi$ and gluing the surface back together.  This does not change the surface, but will affect cycles intersecting the one that was cut.  Specifically, the two Dehn twists on the torus change the one-cycles as \cite{evthesis}
\be
D_a:\begin{matrix}a & \to & a \\ b & \to & b+a\end{matrix}, \hspace{1cm}D_b:\begin{matrix} a & \to & a-b \\ b & \to & b\end{matrix}.
\ee
The effects of each Dehn twist can be absorbed into changes in $\tau$ and $\omega$.  Specifically, after the twist $D_a$ we have a changed $b$-cycle, so we should recalculate
\be
\tau'=\int_{b'}\omega=\int_{b+a}\omega=\int_b\omega+\int_a\omega=\tau+1
\ee
so we see that this twist has had the effect $\tau\to\tau+1$.  After $D_b$ the $a$-cycle is changed.  Now we have
\be
\int_{a'}\omega=\int_{a-b}\omega=1-\tau,
\ee
from which we see that if we take $\omega\to(1-\tau)^{-1}\omega$, we retain the normalization $\int_{a'}\omega'=1$, at the cost of a new $\tau$:
\be
\tau'=\int_b\omega'=(1-\tau)^{-1}\int_b\omega=\frac{\tau}{1-\tau}.
\ee
In summary, the effects of the two twists on the complex structure constant are
\be
D_a:\tau\to\tau+1,\hspace{1cm}D_b:\tau\to\frac{\tau}{1-\tau}.
\ee
These two transformations generate the large diffeomorphisms of the torus.  Note that, under arbitrary composition of $D_a$ and $D_b$, the resulting $\tau$ can always be written in the form $\tau'=(A\tau+B)(C\tau+D)^{-1}$ for integers $A,B,C,D$.  Expressing the Dehn twists in this $\lp\begin{smallmatrix} A & B \\ C & D\end{smallmatrix}\rp$ form gives
\be
D_a=\left(\begin{matrix}1 & 1 \\ 0 & 1\end{matrix}\right), \hspace{1cm}D_b=\left(\begin{matrix}1 & 0 \\ -1 & 1\end{matrix}\right),
\ee
agreeing with one of the SL(2;$\Z$) presentations given in \cite{massuyeau}.

\subsection{Genus Two Mapping Class Group}
\label{g2mapping}

\begin{figure}
	\center
	\begin{tikzpicture}[scale=1]
	\draw (-3.5,0) .. controls (-3.5,2) and (-1.5,2.5) .. (0,2.5);
	\draw[xscale=-1] (-3.5,1) .. controls (-3.5,2) and (-1.5,2.5) .. (0,2.5);
	\draw[rotate=180] (-3.5,1) .. controls (-3.5,2) and (-1.5,2.5) .. (0,2.5);
	\draw[yscale=-1] (-3.5,0) .. controls (-3.5,2) and (-1.5,2.5) .. (0,2.5);
	
	\draw (-2,.2) .. controls (-1.5,-0.3) and (-1,-0.5) .. (0,-.5) .. controls (1,-0.5) and (1.5,-0.3) .. (2,0.2);
	\draw (-1.75,0) .. controls (-1.5,0.3) and (-1,0.5) .. (0,.5) .. controls (1,0.5) and (1.5,0.3) .. (1.75,0);
	
	\draw[thick] (0,0) ellipse (3cm and 2cm);
	\draw[dashed, thick] (0,-1.5) ellipse (0.7cm and 1cm);
	
	\draw[thick] (0,-1.5) [partial ellipse=270:90:0.7cm and 1cm];
	
	\draw (3.5,1) .. controls (3.5,2) and (5.5,2.5) .. (7,2.5);
	\draw[xscale=-1] (-10.5,0) .. controls (-10.5,2) and (-8.5,2.5) .. (-7,2.5);
	\draw[yscale=-1] (3.5,1) .. controls (3.5,2) and (5.5,2.5) .. (7,2.5);
	\draw[yscale=-1] (10.5,0) .. controls (10.5,2) and (8.5,2.5) .. (7,2.5);
	
	\draw (5,.2) .. controls (5.5,-0.3) and (6,-0.5) .. (7,-.5) .. controls (8,-0.5) and (8.5,-0.3) .. (9,0.2);
	\draw (5.25,0) .. controls (5.5,0.3) and (6,0.5) .. (7,.5) .. controls (8,0.5) and (8.5,0.3) .. (8.75,0);
	
	\draw[thick] (7,0) ellipse (3cm and 2cm);
	\draw[dashed, thick] (7,-1.5) ellipse (0.7cm and 1cm);
	
	\draw[thick] (7,-1.5) [partial ellipse=270:90:0.7cm and 1cm];
	
	\draw[dashed, thick] (3.5,0) ellipse (1.6 cm and 0.7cm);
	\draw[thick] (3.5,0) [partial ellipse=0:180:1.6cm and 0.7cm];
	
	\draw[-triangle 90] (-0.7,-1.5) -- +(0,.1);
	\draw[-triangle 90] (0,-2) -- +(.1,0);
	\draw[-triangle 90] (6.3,-1.5) -- +(0,.1);
	\draw[-triangle 90] (7,-2) -- +(.1,0);
	\draw[-triangle 90] (3.5,0.7) -- +(-.1,0);
	
	\node at (0,-1) {$a_1$};
	\node at (0,1.5) {$b_1$};
	\node at (7,-1) {$a_2$};
	\node at (7,1.5) {$b_2$};
	\node at (3.5,0) {$c$};
	\end{tikzpicture} 
	\caption{A genus two Riemann surface, with 5 oriented homotopy one-cycles.}
	\label{genus2figure}
\end{figure}
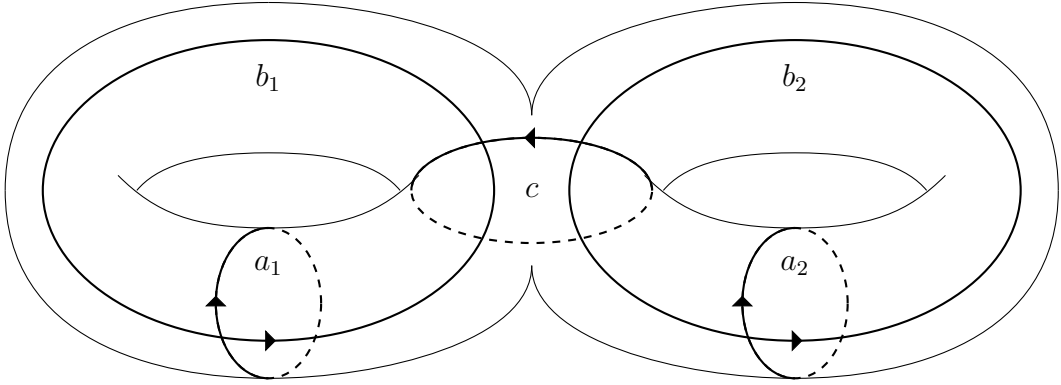

We would like to do something similar for a genus two surface.  We now have two `torus-like' components, which both have their own $a$- and $b$-cycle, giving us four cycles.  In order to generate the entire mapping class group, however, we need to add a fifth cycle \cite{massuyeau}.  Let this cycle run between the two holes, intersecting each $b$-cycle once, and call it $c$ (see figure \ref{genus2figure} for an illustration).  We now have two holomorphic one-forms.  Their normalization along the $a$- and $b$-cycles will be analogous to the torus, but now we have to keep track of their period over the $c$-cycle.  Let
\be
\int_{a_i}\omega_j=\delta_{ij},\hspace{1cm}\int_{b_i}\omega_j=\tau_{ij},\hspace{1cm}\int_c\omega_1=1\hspace{1cm}\int_c\omega_2=-1
\ee
where the result for the $c$-cycle comes from it being homologous to $a_1-a_2$.
The Dehn twists affect the cycles as (leaving other cycles invariant)
\be
D_{a_i}:b_i\to b_i+a_i, \hspace{.1cm}
D_{b_1}:\begin{matrix}a_1 & \to & a_1-b_1 \\ c & \to & c-b_1\end{matrix}, \hspace{.1cm}
D_{b_2}:\begin{matrix}a_2 & \to & a_2-b_2 \\ c & \to & c+b_2\end{matrix}, \hspace{.1cm}
D_{c}:\begin{matrix}b_1 & \to & b_1+c\\b_2 & \to & b_2-c\end{matrix}.
\ee
Running through the same analysis as above, we find that the effect on the periods is given by
\bea
D_{a_1}&:&\left(\begin{matrix}\tau_{11} & \tau_{12} \\ \tau_{21} & \tau_{22}\end{matrix}\right)\to\left(\begin{matrix}\tau_{11}+1 & \tau_{12} \\ \tau_{21} & \tau_{22}\end{matrix}\right)\\
D_{a_2}&:&\left(\begin{matrix}\tau_{11} & \tau_{12} \\ \tau_{21} & \tau_{22}\end{matrix}\right)\to\left(\begin{matrix}\tau_{11} & \tau_{12} \\ \tau_{21} & \tau_{22}+1\end{matrix}\right)\\
D_{b_1}&:&\left(\begin{matrix}\tau_{11} & \tau_{12} \\ \tau_{21} & \tau_{22}\end{matrix}\right)\to\frac{1}{1-\tau_{11}}\left(\begin{matrix}\tau_{11} & \tau_{12} \\ \tau_{21} & \tau_{22}-\text{det }\tau\end{matrix}\right)\\
D_{b_2}&:&\left(\begin{matrix}\tau_{11} & \tau_{12} \\ \tau_{21} & \tau_{22}\end{matrix}\right)\to\frac{1}{1-\tau_{22}}\left(\begin{matrix}\tau_{11}-\text{det }\tau & \tau_{12} \\ \tau_{21} & \tau_{22}\end{matrix}\right)\\
D_{c}&:&\left(\begin{matrix}\tau_{11} & \tau_{12} \\ \tau_{21} & \tau_{22}\end{matrix}\right)\to\left(\begin{matrix}\tau_{11}+1 & \tau_{12}-1 \\ \tau_{21}-1 & \tau_{22}+1\end{matrix}\right).
\eea
These 5 operators have realizations given in (\ref{g2mcgpres1}-\ref{g2mcgpres2}), again akin to the torus, as matrices which should constitute a presentation of the genus two modular group Sp$(4;\Z)$.

%We see that the Dehn twists $D_{a_i}$ and $D_{b_i}$ generate a Sp$(4;\Z)$ action on the period matrix.  Ignoring the $c$-cycle, coincidentally, amounts to treating the surface as a product of tori, i.e. a 4-torus.  The quantities out of which we built partition functions in previous sections were defined, via the Abel map, on the Jacobian associated with our genus two surface, which is a 4-torus; it should come as no surprise that (at least on the partition function level) the CFT is happy to `ignore' the $c$-cycle.  When we consider the insertion of group elements as part of the orbifold construction, however, we may be forced to consider transformation under the full mapping class group.

\subsection{Orbifold Partial Traces}

%In constructing a (torus) orbifold theory we assign a group element to each homology cycle.  For instance, to build the partial trace $Z_{h,g}$ we assign $h\in G$ to the $a$-cycle and $g\in G$ to the $b$-cycle.  In the path integral formulation this is equivalent to specifying boundary conditions along these cycles. At the level of the partition function, $Z_{1,g}$ can be represented by inserting (a representation of) $g$ into the trace over states, while $Z_{h,1}$ is given by tracing over a modified Hilbert space $\H_h$.

What is the effect of a torus Dehn twist on a partial twist $Z_{k,g}$, as defined in (\ref{partialtrace})?   We established that $D_a$ takes the $b$-cycle to $b+a$, so after the twist going around $b$ would implement both $g$ and $k$, so we expect $D_a:Z_{k,g}\to Z_{k,gk}$.  Similarly, $D_b$ takes $a$ to $a-b$, so we should have $D_b:Z_{k,g}\to Z_{kg^{-1},g}$.  Composition of these operations can then generate partial traces labeled by arbitrary combinations of powers of $g$ and $k$.  Since we have explicit forms of $D_a$ and $D_b$ in terms of their action on the complex structure constant, we can construct any such partial trace in terms of untwisted sector partition functions with varying arguments in $\tau$.

Turning back to genus two, we should be assigning a group element to all four cycles now, so our partial traces should take the form $Z_{k,l;m,n}$.  Similarly to the torus, the action of the Dehn twists on these objects should be
\bea
D_{a_1}&:&Z_{k,l;m,n}\to Z_{k,l;mk,n} \\
D_{a_2}&:&Z_{k,l;m,n}\to Z_{k,l;m,nl} \\
D_{b_1}&:&Z_{k,l;m,n}\to Z_{km^{-1},l;m,n} \\
D_{b_2}&:&Z_{k,l;m,n}\to Z_{k,ln^{-1};m,n} \\
D_{c}&:&Z_{k,l;m,n}\to Z_{k,l;mkl^{-1},nk^{-1}l}.
\eea

We can see that, since the $a$ and $b$ twists generate $\text{SL}(2;\Z)\times\text{SL}(2;\Z)$, in any situation where $\text{SL}(2,\Z)$ generated the full partition function at genus one, we can get the full genus two partition function without $D_c$.  Recall that, in the case of an orbifold by $\Z$, we can label our group elements by integers.  At genus one, we can reach an arbitrary partial trace $Z_{m,n}$ from the untwisted sector through the modular transformation
\be
\label{modtransg1}
Z_{m,n}(\tau,\btau)=Z_{0,r}(r\frac{a\tau+b}{n-m\tau},r\frac{a\btau+b}{n-m\btau})
\ee
where we have chosen $a$ and $b$ such that $an+bm=r=\text{gcd}(m,n)$.  From these conditions one sees that the matrix
\be
\lp\begin{matrix}a & b \\ -m/r & n/r\end{matrix}\rp
\ee
is in $\text{SL}(2;\Z)$.  At genus two, in order to reach an arbitrary partial trace $Z_{m_1,m_2;n_1,n_2}$, we make the analogous $\text{SL}(2;\Z)\times\text{SL}(2;\Z)$ transformation
\be
\label{sp4zcyclic}
\lp\begin{matrix}a_1 & 0 & b_1 & 0 \\ 0 & a_2 & 0 & b_2 \\ -m_1/r_1 & 0 & n_1/r_1 & 0 \\ 0 & -m_2/r_2 & 0 & n_2/r_2\end{matrix}\rp
\ee
where, of course, we have picked $a_1n_1+b_1m_1=r_1=\text{gcd}(m_1,n_1)$ and $a_2n_2+b_2m_2=r_2=\text{gcd}(m_2,n_2)$.  Our partial trace can then be calculated as (antiholomorphic arguments omitted to save space)
\begin{align}
\label{modtransg2}
&Z_{m_1,m_2;n_1,n_2}\left(\left[\begin{matrix}\tau_{11} & \tau_{12} \\ \tau_{12} & \tau_{22}\end{matrix}\right]\right)\\\notag
&=Z_{0,0;r_1,r_2}\left(\lp\left[\begin{matrix}a_1 & 0 \\ 0 & a_2\end{matrix}\right]\left[\begin{matrix}\tau_{11} & \tau_{12} \\ \tau_{12} & \tau_{22}\end{matrix}\right]+\left[\begin{matrix}b_1 & 0 \\ 0 & b_2\end{matrix}\right]\rp\lp\left[\begin{matrix}-\frac{m_1}{r_1} & 0 \\ 0 & -\frac{m_2}{r_2}\end{matrix}\right]\left[\begin{matrix}\tau_{11} & \tau_{12} \\ \tau_{12} & \tau_{22}\end{matrix}\right]+\left[\begin{matrix} \frac{n_1}{r_1} & 0 \\ 0 & \frac{n_2}{r_2}\end{matrix}\right]\rp^{-1}\right).
\end{align}
The full partition function is then obtained by summing over $m_1,m_2,n_1,n_2$.  We expect that different choices of the pairs $a_1,b_1$ and $a_2,b_2$ will not affect the form of the partial traces; we will see this explicitly for theories with continuous symmetries in section \ref{flavored}.

\subsection{Test Case: $\Z_2\times\Z_2$}
\label{z2z2}

There are, of course, situations where we are unable to generate the entire genus one partition function starting from the untwisted sector.  Perhaps the simplest such case is an orbifold by $\Z_2\times\Z_2$.  In that case, we are able to put constraints on the torus partition function just by examining the structure of orbits at genus two.  Let's begin by establishing notation for the elements of $\Z_2\times\Z_2$:
\be
(0,0)\equiv1\hspace{.5cm}(1,0)\equiv x\hspace{.5cm}(0,1)\equiv y\hspace{.5cm}(1,1)\equiv z.
\ee
We'll quickly review the genus one case.  Beginning in the untwisted sector, we have partial traces $Z_{11},Z_{1x},Z_{1 y},Z_{1z}$.  SL$(2;\Z)$ modular orbits generate 6 more: $Z_{x1},Z_{xx},Z_{y1},Z_{yy},Z_{z1},Z_{zz}$.  These 10 are all that can be reached from the untwisted sector via modular orbits.  There is one additional orbit, consisting of the remaining 6 partial traces, which must be added in by hand: $Z_{xy},Z_{xz},Z_{yx},Z_{yz},Z_{zx},Z_{zy}$.

Moving back to genus two, do we see similar behavior?  Note that, under a separating degeneration, we expect the partial traces to behave as
\be
Z_{a_1a_2b_1b_2}\to Z_{a_1b_1}\cdot Z_{a_2b_2}.
\ee
We can begin to map things out starting from the untwisted sector and noting that the $a$ and $b$ Dehn twists act exactly as they would in the genus one case.  Then the indices $(a_1,b_1)$ and $(a_2,b_2)$ can separately take on all ten combinations reachable from the genus one untwisted sector.  It's up to $D_c$ to give us new combinations, which it does.  Take for instance $D_cZ_{xz11}=Z_{xzyy}$.  This would degenerate to $Z_{xy}\cdot Z_{zy}$, both of which are inaccessible from the genus one untwisted sector.  However, one can check that traces which degenerate to one accessible and one inaccessible genus one trace \textit{cannot} be generated this way (e.g. there's no way to reach $Z_{xyzy}$ with Dehn twists starting from the untwisted sector).  

Schematically, from taking modular orbits of untwisted sector partial traces and then degenerating, we get
%degeneration of a genus two $\Z_2\times\Z_2$ orbifold, we would get
\be
\label{eq:Z2Z2Degeneration}
\text{Accessible}_{T^2_1}\cdot\text{Accessible}_{T^2_2}+\text{Inaccessible}_{T^2_1}\cdot\text{Inaccessible}_{T^2_2}.
\ee
On the other hand, we expect the full genus two partition function should degenerate to a product of genus one partition functions, for which (\ref{eq:Z2Z2Degeneration}) is missing the cross-terms.
However, even without the cross-terms, we can determine the full partition function on the torus up to a relative sign: $Z_{g=1}\simeq(\text{Accessible}\pm\text{Inaccessible})$.  This ambiguity is exactly the choice of discrete torsion.

\section{Flavored Partition Functions}
\label{flavored}

\subsection{Review of Genus One}

At genus one, it is sometimes useful to define a \textit{flavored} partition function in which we keep track of additional quantum numbers.  For a CFT with holomorphic (antiholomorphic) currents $J_L$ $(J_R)$, the flavored partition function is
\be
Z^f(\tau,\bar{\tau},z_L,z_R)=\text{Tr}\left[q^{L_0-\frac{c}{24}}\bar{q}^{\bar{L}_0-\frac{\bar{c}}{24}}e^{2\pi iz_LJ_L}e^{-2\pi iz_RJ_R}\right]
\ee
and behaves under modular transformations as
\be
\label{zftrans}
Z^f\left(\frac{a\tau+b}{c\tau+d},\frac{a\bar{\tau}+b}{c\bar{\tau}+d},z_L,z_R\right)=e^{\pi ik\left(c(c\tau+d)z_L^2-c(c\bar{\tau}+d)z_R^2\right)}Z^f(\tau,\bar{\tau},(c\tau+d)z_L,(c\bar{\tau}+d)z_R).
\ee
Such a setup is particularly suited for the orbifold procedure \cite{robbins2019orbifolds}.  For orbifold group $G=\Z$ (or possibly $\Z_N$), combining (\ref{modtransg1}) with (\ref{zftrans}) gives the partial traces in terms of the flavored partition function as
\be
\label{zmng1}
Z_{m,n}(\tau)=e^{\pi ik\left(m^2(\tau\alpha_L^2-\bar{\tau}\alpha_R^2)-mn(\alpha_L^2-\alpha_R^2)\right)}Z^f(\tau,(n-m\tau)\alpha_L,(n-m\bar{\tau})\alpha_R)
\ee
where we've chosen the $\alpha$ such that $e^{2\pi i\alpha J}\in G$, so that flavoring the partition function corresponds to inserting a group element.  Writing the partition function as a sum over the weights of CFT states allows us to isolate the sum over $n$, which has the form of a projector.  Its role is to regulate which states show up in the twisted sectors such that modular invariance is preserved.

For example, in the case of the free boson, an obvious choice of conserved current is the $U(1)\times U(1)$ generated by $p_L$ and $p_R$.  Given this choice, the partition function in the $m$-twisted sector takes the form
\be
\label{mtwistg1}
Z_m=|\eta(\tau)|^{-2}\sum q^{\frac{1}{4}(p_L-2m\alpha_L)^2}\bar{q}^{\frac{1}{4}(p_R-2m\alpha_R)^2}.
\ee
where the sum is over states on the momentum lattice allowed by the projection constraint
\be
\label{genus1proj}
\alpha_Lp_L-\alpha_Rp_R-m(\alpha_L^2-\alpha_R^2)\in\Z.
\ee

Picking $\alpha_L$ and $\alpha_R$ appropriately, one can straightforwardly construct the partition functions of (asymmetric or symmetric) orbifold theories, though there's no guarantee the resulting partition function will differ from the parent one (which simply means the orbifold was not consistent with modular invariance).

\subsection{Higher Genus Flavored Orbifolds}

On the torus we defined the flavored partition function by inserting terms in the trace over states, but for the boson one can equally well express the partition function as a sum over the momentum lattice, which will readily generalize to higher genus.  Defining
\begin{multline}
\label{gen2flav}
Z^f_{g=2}(\tau,\bar{\tau},\alpha_L,\alpha_R)\equiv H_2(\tau)\sum_{p_L,p_R\in\Gamma_2}\exp{[2\pi i \alpha_L\cdot p_L]}\exp{[-2\pi i \alpha_R\cdot p_R]}\\
\times\exp{\left[\frac{2\pi i}{4}(p_L\cdot\tau\cdot p_L-p_R\cdot\bar{\tau}\cdot p_R)\right]},
\end{multline}
we're at least guaranteed analogous transformation properties to (\ref{zftrans}) under SL$(2;\Z)\times$ SL$(2;\Z)$.  Then, making the choice (\ref{sp4zcyclic}) of modular transformation, we can calculate the partial traces of a $\Z$ orbifold as in (\ref{modtransg2}).

Running through the same calculation that led to (\ref{zmng1}), we arrive (unsurprisingly, but perhaps reassuringly) at a result that straightforwardly generalizes (\ref{mtwistg1}):
\begin{multline}
\label{genus2orbifold}
Z_{m_1,m_2}=H_2(\tau)\sum\text{exp }\ls\frac{2\pi i}{4}(p_L-2m\alpha_L)\cdot\tau\cdot(p_L-2m\alpha_L)\rs\\
\times\text{exp}\ls -\frac{2\pi i}{4}(p_R-2m\alpha_R)\cdot\bar{\tau}\cdot(p_R-2m\alpha_R)\rs
\end{multline}
where $\alpha$, $p$ and $m$ are now two-component vectors (here $m\alpha_L$ should be understood as a vector $(m_1\alpha_{L_1},m_2\alpha_{L_2})$, likewise for $m\alpha_R$.) and the sum is over $p_L,p_R\in\Gamma_2$ subject to the individual projection constraints
\be
\label{genus2proj1}
\alpha_{L_1}p_{L_1}-\alpha_{R_1}p_{R_1}-m_1(\alpha_{L_1}^2-\alpha_{R_1}^2)\in\Z
\ee
\be
\label{genus2proj2}
\alpha_{L_2}p_{L_2}-\alpha_{R_2}p_{R_2}-m_2(\alpha_{L_2}^2-\alpha_{R_2}^2)\in\Z.
\ee

We need to determine what the possible consistent choices for $\alpha_{L_1},\alpha_{L_2},\alpha_{R_1},\alpha_{R_2}$ are.  When $\alpha_{L_1}=\alpha_{L_2}$ and $\alpha_{R_1}=\alpha_{R_2}$, this expression clearly has the expected leading order separating degeneration behavior, giving a copy of (\ref{mtwistg1}) on each torus (along with the appropriate projection).

What happens if we pick $\alpha_{L_1/R_1}\neq\alpha_{L_2/R_2}$?  Since our orbifold was built through SL$(2;\Z)\times$ SL$(2;\Z)$ orbits, it would appear that this choice could put elements of different groups on the $(a_1,b_1)$ cycles than the $(a_2,b_2)$ cycles.  But we have to remember that Dehn twists around the $c$ cycle exist, and need to be taken into account to ensure invariance under the full modular group.  Abbreviating $D_a\equiv D_{a_1}D_{a_2}$ and $D_b\equiv D_{b_1}D_{b_2}$ we can, in the case of abelian groups for example, build actions such as
\be
D_bD^2_aD_b^2D_aD_cD_bD_aD_cD_bZ_{0,0,g,k}(\tau)=Z_{0,0,k,g}(\tau),
\ee
which acts trivially on the period matrix but swaps the group elements on the $b_1$ and $b_2$ cycles.  In general, if we try to pick $\alpha_{L_1/R_1}\neq\alpha_{L_2/R_2}$ such that $\alpha_{L_1/R_1}$ leads to a $\Z_{N_1}$ action and $\alpha_{L_2/R_2}$ to $\Z_{N_2}$, the full set of modular orbits will be equivalent to a $\Z_{\text{lcm}(N_1,N_2)}$ orbifold with $\alpha_{L_1/R_1}=\alpha_{L_2/R_2}$.  Modularity effectively forces us to choose $\alpha_{L_1/R_1}=\alpha_{L_2/R_2}$.

\subsection{Flavored Orbifold Correlation Functions}

Now that we've computed the orbifold theory's partition function, we can extract CFT data.  As we did with the Ising model in section \ref{isingcorr}, we'll apply the degeneration procedure to our boson partition functions.  We begin with the usual compact boson (\ref{genusgz}) -- modifying the results in the orbifold theory will be straightforward.

Again we begin with the separating degeneration.  For a generic radius, from (\ref{separatingschematic}) we expect the $|t|^2$ term in this expansion to have as its coefficient the square of the $2\partial\varphi\bar{\partial}\varphi$ torus one-point function (the factor of two is chosen so that, in our conventions, the operator has unit normalized two-point function on the sphere).  Using the period matrix (\ref{pmsepdegen}) and taking both a $t$ and $\bar{t}$ derivative, we find the coefficient of the $|t|^2$ term to be\footnote{This analysis assumes that $H_2(\tau)$ has no $|t|^2$ term in its degeneration series.}
\be
4\pi^4|\eta(\tau_1)|^{-2}|\eta(\tau_2)|^{-2}\sum_{\substack{p_{L_1},p_{R_1}\in\Gamma_1 \\ p_{L_2},p_{R_2}\in\Gamma_1}}p_{L_1}p_{R_1}p_{L_2}p_{R_2}\exp{\left[\frac{2\pi i}{4}(p_{L_1}^2\tau_1+p_{L_2}^2\tau_2-p_{R_1}^2\tau_1-p_{R_2}^2\tau_2)\right]},
\ee
from which we identify
\be
\label{dphidphionept}
\braket{2\partial\varphi\bar{\partial}\varphi(0)}=\pm2\pi^2|\eta(\tau)|^{-2}\sum_{p_L,p_R\in\Gamma_1}p_Lp_R\exp{\left[\frac{2\pi i}{4}(p_L^2\tau-p_R^2\bar{\tau})\right]}.
\ee
In light of the result for the orbifold partition function (\ref{genus2orbifold}), in a flavored orbifold theory this correlation function should take the form
\begin{multline}
\braket{2\partial\varphi\bar{\partial}\varphi(0)}_{\text{orb.}}=\pm 2\pi^2|\eta(\tau)|^{-2}\sum_{\substack{p_L,p_R\in\Gamma_1 \\ m\in\Z_N \\ \text{conditional to (\ref{genus1proj})}}}(p_L-2m\alpha_L)(p_R-2m\alpha_R) \\
\times\exp{\left[\frac{2\pi i}{4}((p_L-2m\alpha_L)^2\tau-(p_R-2m\alpha_R)^2\bar{\tau})\right]}.
\end{multline}

Turning to the separating degeneration, we now use the period matrix (\ref{nonseparatingschematic}).  The logarithms from the off-diagonal terms give us our $t$ and $\bar{t}$ terms, which appear here raised to powers of the momentum running in the degenerating cycle.  Specifically, for the coefficient of $t^{k_L^2/4}\bar{t}^{k_R^2/4}$ we find
\be
\label{boson2point}
2[E(z,0)]^{-\frac{k_L^2}{2}}[E(\bar{z},0)]^{-\frac{k_R^2}{2}}|\eta(\tau)|^{-2}\sum_{p_L,p_R\in\Gamma_1}\exp{\left[\frac{2\pi i}{4}(\tau p_L^2-\bar{\tau}p_R^2+2zp_Lk_L-2\bar{z}p_Rk_R)\right]}.
\ee

Recall that in the free boson theory we have vertex operators of the form
\be
\Or_{k_L,k_R}=\sqrt{2}\cos{(k_L\varphi_L+k_R\varphi_R)},\hspace{.5cm}\Or'_{k_L,k_R}=\sqrt{2}\sin{(k_L\varphi_L+k_R\varphi_R)}.
\ee
Both of these operators have weights $h=k_L^2/4,\bar{h}=k_R^2/4$ and so are degenerate.  In taking the non-separating degeneration limit, we are finding the specific two-point functions
\be
\braket{\Or_{k_L,k_R}(z)\Or_{k_L,k_R}(0)}+\braket{\Or'_{k_L,k_R}(z)\Or'_{k_L,k_R}(0)},
\ee
as it is this combination which takes the form of $z^{-k_L^2/2}\bar{z}^{-k_R^2/2}$ times a series in integer powers of $z$ and $\bar{z}$.

As in the Ising model case, we can expand these two-point functions to find CFT data.  Noting that, to leading order, $E(z,0) \sim z$ and differentiating the lattice sum once in $z$ and once in $\bar{z}$ yields
\be
k_Lk_Rz^{1-k_L^2/2}\bar{z}^{1-k_R^2/2}\cdot 2\pi^2|\eta(\tau)|^{-2}\sum_{p_L,p_R\in\Gamma_1}p_Lp_R\exp{\left[\frac{2\pi i}{4}(\tau p_L^2-\bar{\tau}p_R^2)\right]}.
\ee
Comparing this with (\ref{dphidphionept}), we can pick out
\be
\lambda_{\Or_{k_L,k_R}\Or_{k_L,k_R}2\partial\varphi\bar{\partial}\varphi}+\lambda_{\Or'_{k_L,k_R}\Or'_{k_L,k_R}2\partial\varphi\bar{\partial}\varphi}=\pm k_Lk_R.
\ee
In fact, a direct computation of OPEs reveals that the right-hand side is $-k_Lk_R$, so our procedure seems to be consistent.

Again, the analogous computation for the orbifold proceeds similarly, and we find in that case the same result, but with $k_L\to k_L-2m\alpha_L$, $k_R\to k_R-2m\alpha_R$ and only holding when the projection constraint (\ref{genus1proj}) is satisfied, i.e. only for $(k_L,k_R,m)$ satisfying $\alpha_Lk_L-\alpha_Rk_R-m(\alpha_L^2-\alpha_R^2)\in\Z.$

\subsection{Correlation Function Generalities}
\label{subsec:CFG}

Finally we present some thoughts and observations on calculating correlation functions and OPE coefficients in more general theories.  The methods of this paper provide two slightly different routes to such results.  First, as in the preceding examples, one could calculate the theory's partition function(s) at higher genus.  Degeneration allows for calculation of multi-point functions, then expansion in $z$ and comparison to relations such as (\ref{opeexpansion}) yields OPE coefficients.  In the case of theories with continuous symmetries we have presented means for simplifying the calculation.

The second method would be to begin with the parent theory genus one correlation functions $\braket{\Or(z)}$ (written here as one-point functions, but in general could be multi-point) and compute their equivalents of partial traces, defined in terms of a path integral in (\ref{ptor}).  Once one has the objects $\braket{\Or}_{k,g}$, computing correlation functions in the orbifold theory is straightforward.  For each $k\in G$ one can form
\be
\braket{\Or}^{\text{orb.}}_{k}=\frac{1}{|G|}\sum_{g\in G}\braket{\Or}_{k,g}.
\ee
For $k$ the identity this simply reproduces the parent theory correlation functions that were invariant under the group action.  For other values of $k$ we produce the twisted sector correlation functions.  Once again, expanding these in $z$ allows one to identify OPE coefficients.

Both methods have advantages and disadvantages.  Ideally one might hope to produce partition functions at arbitrary genus and extract the desired information that way.  However, explicit computations at arbitrary genus quickly become difficult to intractable, so this may not be feasible.  Additionally, higher order correlation functions will require expanding the period matrix (and any other relevant quantities) to higher orders in the degeneration parameter, which may also become unwieldy.  Working directly with the genus one correlation functions means that one can entirely avoid working on higher genus surfaces, and can begin directly from the desired order of multi-point function.  The downside to this method is that it may not be clear how to evaluate $\braket{\Or}_{k,g}$, even knowing the parent theory correlators.  Perhaps the best use case of the correlation function partial trace method would be a scenario in which one can express the genus one correlation functions in terms of a sum over states, but has no access to higher genus partition functions.  Then, just as for the partition function, one should be able to evaluate $\braket{\Or}_{1,g}$ for the untwisted sector (by inserting a representation of the group acting on the states), and fill in the twisted sectors by modular orbits.  This approach has the potential flaw that there can be orbits which do not involve the untwisted sector -- in \cite{robbins2019orbifolds} we presented one workaround for this issue when calculating partition functions of orbifolds by groups that are solvable, in the form of an iterated orbifold procedure.  We expect such a method to work as well for correlation functions.

There should also be a third, more direct route to make contact with the OPE coefficients.  In yet another degeneration limit (a refinement of the non-separating degeneration), we can view the genus two surface as a pair of spheres connected by three long thin tubes.  By inserting complete sets of states in each tube, we can relate the genus two partition function to sums of squares of sphere three-point functions~\cite{Cardy:2017qhl}, weighted by particular powers of degeneration parameters depending on the operators involved.  The sphere three-point functions are in turn directly related to OPE coefficients.  Although we understand how this works schematically, and can verify some relations at low orders in conformal weights, fixing all of the details of this approach is work in progress.

\section{Conclusion}
\label{conclusion}

At genus one we have a very nice general expression for a theory's partition function (at least in the case of a discrete, diagonalizable spectrum) given by
\be
\label{g1partitiontrace}
Z(\tau)=\Tr_{h,\bar{h}}{\left[\exp{[2\pi i\tau(h-\frac{c}{24})]}\exp{[-2\pi i\bar{\tau}(\bar{h}-\frac{\bar{c}}{24})]}\right]},
\ee
which cleanly encodes CFT data given by the spectrum $(h,\bar{h})$ as a function of the surface geometry, captured in the complex structure constant $\tau$.  At higher genus there is, in general, no equally nice expression; we might have expected this, since we are now necessarily encoding more information than just the spectrum.  It is not unreasonable to wonder, given a theory to start with, how much we need to know or specify to construct an orbifold.  We have argued here that so long as one knows the partition functions (at various genera) of the parent theory and understands how the orbifold group modifies those (in the form of untwisted sector partial traces), modular invariance will dictate the rest.

In the specific case of (theories which can be cast as) free bosons, we have the notion of a momentum lattice, which allows us to cast (\ref{g1partitiontrace}) in the form (\ref{genusgz}), which \textit{does} generalize quite readily to higher genus.  This provides a rich testing ground for our ideas, as the technology of flavored partition functions allows us to demonstrate our proposal in a fully explicit nature.  Orbifolds by arbitrary cyclic actions (both symmetric and asymmetric) built out of momenta have at genus two the partition function (\ref{genus2orbifold}), the form of which holds for higher genera as well.

The analysis at higher genus comes full circle in addressing some of the potential concerns laid out with the genus one version of this procedure in \cite{robbins2019orbifolds}.  One of the potentially glaring issues with modular orbits is that not all orbits can be reached from the untwisted sector through modular transformations, threatening to leave our procedure incomplete.  As we saw explicitly in section \ref{z2z2}, these disconnected orbits will make themselves present in higher genus partition functions, so the process of degeneration can be used to fill out full genus one partition functions.  Further, combined with genus two modular invariance, these disconnected orbits should show up with an appropriately constrained choice of phase (which is, though we did not show it here, dictated by H$^2(G,U(1))$).  We have focused our explicit examples on genus two in this paper, partly because it is the simplest example past the torus, but also because it is known that modular invariance at genus one and two is sufficient to fully determine the constraints of discrete torsion on how orbits combine \cite{VafaTorsion}.  This parallels another solution to this issue which can be implemented purely at genus one in which an orbifold by a solvable group is built up in an iterated fashion.  Here the choice of discrete torsion appears as a choice of how successive actions behave in the twisted sectors of their predecessors.

There was also the potential that using a pure modular orbits method, we may have ended up computing modular invariant objects which had no sensible interpretation as the partition function of any CFT.  A preliminary check on this was that our expressions led to multiplicities that were non-negative integers.  Higher genus calculations go further towards validating our methods -- now we have seen that the expressions we obtain behave in the expected way under worldsheet degeneration.  Furthermore, we are able to compute sensible correlation functions, and in all cases where we were able to compare to alternative calculational methods our results were found to match.

There are several directions that could be followed from here.  One of our original motivations for understanding, in detail, the precise connections between genus two partition functions and the data (spectrum and OPE coefficients, or equivalently, correlation functions of local operators) was to be able to apply the philosophy and methods of the modular bootstrap program to genus two.  Some work in this direction has been done~\cite{Cardy:2017qhl,Keller:2017iql,Cho:2017fzo}, and we would like to systematize this approach.

Our approach to orbifolds also opens up the possibility of computing OPE coefficients in situations where neither the orbifold nor the parent theory have a known free field realization.  We have given a prescription for constructing the genus two partition functions even in such cases, as long as the genus two partition function (and related objects with insertions of group elements) are known for the parent theory.  As discussed in section \ref{subsec:CFG}, we can extract lower genus correlation functions and OPE coefficients from there (perhaps up to some extra discrete data).  This could be relevant for model building (where, for instance certain OPE coefficients translate to physical quantities such as Yukawa couplings).

Finally, an interesting direction to move would be to combine this work with the idea of conformal interfaces, topological or otherwise~\cite{Bachas:2001vj,Bachas:2007td,Gaiotto:2012np,Quella:2002ct,Brunner:2003dc,Brunner:2007qu,Konechny:2015qla,Graham:2003nc,Fuchs:2015ska,Bachas:2013ora,Becker:2017zai}.  These defects can be used to formulate many aspects of 2D CFTs and the RG flows between them, and little work has been done on higher genus aspects of this formulation.

%Say something about what to do next with our methods, including bootstrappy stuff.

\section*{Acknowledgments}

The authors would like to thank O.~Lunin and the other members of the University at Albany string group for helpful conversations.  This material is based upon work supported by the National Science Foundation under Grant No.\ PHY-1820867.

\appendix

\section{Ising Orbifold at Genus Two}
\label{isingorbg2}

To compute the Ising model orbifold of section \ref{isingorbifold}, we will use a set of coordinates on moduli space given by
\be
q_1=e^{2\pi i(\tau_{11}-\tau_{12})},\qquad q_2=e^{2\pi i(\tau_{22}-\tau_{!2})},\qquad q_3=e^{2\pi i\tau_{12}}.
\ee
These coordinates are naturally adapted to the picture of the Riemann surface as a pair of three-punctured spheres connected by tubes anchored at the punctures.  Each coordinate $q_i$ describes the moduli (length and twist) of one of the three tubes.

In terms of the $q_i$ instead of $\tau$,
\be
\theta\ls\begin{matrix}\al_1 & \al_2 \\ \beta_1 & \beta_2\end{matrix}\rs(z|q)=\sum_{n\in\Z^2}q_1^{\hlf\lp n_1+\al_1\rp^2}q_2^{\hlf\lp n_2+\al_2\rp^2}q_3^{\hlf\lp n_1+n_2+\al_1+\al_2\rp^2}e^{2\pi i\lp n+\al\rp^T\cdot\lp z+\beta\rp}.
\ee
Under a modular transformation given by an $\operatorname{Sp}(4,\Z)$ matrix $\lp\begin{smallmatrix}A & B \\ C & D\end{smallmatrix}\rp$, $\tau$ and $z$ transform as
\be
\tau\longrightarrow\widetilde{\tau}=\lp A\cdot\tau+B\rp\cdot\lp C\cdot\tau+D\rp^{-1},\qquad z\longrightarrow\widetilde{z}=\lp\tau\cdot C^T+D^T\rp^{-1}\cdot z.
\ee
If we define
\be
\al'=D\cdot\al-C\cdot\beta+\hlf\operatorname{diag}(CD^T),\quad\beta'=-B\cdot\al+A\cdot\beta+\hlf\operatorname{diag}(AB^T),
\ee
then the theta functions transform as
\be
\theta\ls\begin{matrix}\al' \\ \beta'\end{matrix}\rs(\widetilde{z}|\widetilde{\tau})=e^{i\phi}\det\lp C\cdot\tau+D\rp^{1/2} e^{i\pi z\cdot\lp C\cdot\tau+D\rp^{-1}\cdot C\cdot z}\theta\ls\begin{matrix}\al \\ \beta\end{matrix}\rs(z|\tau).
\ee
Here $\phi$ is a phase that we won't need to worry about.  Finally, if we omit the $z$ argument of the theta function, it should be assumed that we take $z=0$.

According to \cite{Behera:1989gg}, the genus two partition function for the Ising model has the form~(\ref{zising})
\begin{multline}
Z^{(2)}=H_2(\tau)\left\{\left|\theta\ls\begin{matrix}0 & 0 \\ 0 & 0\end{matrix}\rs(\tau)\right|+\left|\theta\ls\begin{matrix}0 & 0 \\ 0 & \hlf\end{matrix}\rs(\tau)\right|+\left|\theta\ls\begin{matrix}0 & 0 \\ \hlf & 0\end{matrix}\rs(\tau)\right|+\left|\theta\ls\begin{matrix}0 & 0 \\ \hlf & \hlf\end{matrix}\rs(\tau)\right|\right.\\
\left. +\left|\theta\ls\begin{matrix}\hlf & 0 \\ 0 & 0\end{matrix}\rs(\tau)\right|+\left|\theta\ls\begin{matrix}\hlf & 0 \\ 0 & \hlf\end{matrix}\rs(\tau)\right|+\left|\theta\ls\begin{matrix}0 & \hlf \\ 0 & 0\end{matrix}\rs(\tau)\right|+\left|\theta\ls\begin{matrix}0 & \hlf \\ \hlf & 0\end{matrix}\rs(\tau)\right|\right.\\
\left. +\left|\theta\ls\begin{matrix}\hlf & \hlf \\ 0 & 0\end{matrix}\rs(\tau)\right|+\left|\theta\ls\begin{matrix}\hlf & \hlf \\ \hlf & \hlf\end{matrix}\rs(\tau)\right|\right\}
\end{multline}
We do not need to worry about the details of $H_2(\tau)$ except that under modular transformations it transforms as
\be
H_2(\widetilde{\tau})=\left|\det\lp C\cdot\tau+D\rp\right|^{-1/2}H_2(\tau),
\ee
which makes $Z^{(2)}$ modular invariant.

To get a better sense of the different pieces, let's look at them in the $q_i$ variables,
\bea
\theta\ls\begin{matrix}0 & 0 \\ 0 & 0\end{matrix}\rs(\tau) &=& \sum_{n,m\in\Z}q_1^{\hlf n^2}q_2^{\hlf m^2}q_3^{\hlf\lp n+m\rp^2},\\
\theta\ls\begin{matrix}0 & 0 \\ 0 & \hlf\end{matrix}\rs(\tau) &=& \sum_{n,m\in\Z}\lp -1\rp^mq_1^{\hlf n^2}q_2^{\hlf m^2}q_3^{\hlf\lp n+m\rp^2},\\
\theta\ls\begin{matrix}0 & 0 \\ \hlf & 0\end{matrix}\rs(\tau) &=& \sum_{n,m\in\Z}\lp -1\rp^nq_1^{\hlf n^2}q_2^{\hlf m^2}q_3^{\hlf\lp n+m\rp^2},\\
\theta\ls\begin{matrix}0 & 0 \\ \hlf & \hlf\end{matrix}\rs(\tau) &=& \sum_{n,m\in\Z}\lp -1\rp^{n+m}q_1^{\hlf n^2}q_2^{\hlf m^2}q_3^{\hlf\lp n+m\rp^2}, \\
\theta\ls\begin{matrix}\hlf & 0 \\ 0 & 0\end{matrix}\rs(\tau) &=& \sum_{n,m\in\Z}q_1^{\hlf\lp n+\hlf\rp^2}q_2^{\hlf m^2}q_3^{\hlf\lp n+m+\hlf\rp^2},\\
\theta\ls\begin{matrix}\hlf & 0 \\ 0 & \hlf\end{matrix}\rs(\tau) &=& \sum_{n,m\in\Z}\lp -1\rp^mq_1^{\hlf\lp n+\hlf\rp^2}q_2^{\hlf m^2}q_3^{\hlf\lp n+m+\hlf\rp^2},\\
\theta\ls\begin{matrix}0 & \hlf \\ 0 & 0\end{matrix}\rs(\tau) &=& \sum_{n,m\in\Z}q_1^{\hlf n^2}q_2^{\hlf\lp m+\hlf\rp^2}q_3^{\hlf\lp n+m+\hlf\rp^2},\\
\theta\ls\begin{matrix}0 & \hlf \\ \hlf & 0\end{matrix}\rs(\tau) &=& \sum_{n,m\in\Z}\lp -1\rp^nq_1^{\hlf n^2}q_2^{\hlf\lp m+\hlf\rp^2}q_3^{\hlf\lp n+m+\hlf\rp^2},\\
\theta\ls\begin{matrix}\hlf & \hlf \\ 0 & 0\end{matrix}\rs(\tau) &=& \sum_{n,m\in\Z}q_1^{\hlf\lp n+\hlf\rp^2}q_2^{\hlf\lp m+\hlf\rp^2}q_3^{\hlf\lp n+m+1\rp^2},\\
\theta\ls\begin{matrix}\hlf & \hlf \\ \hlf & \hlf\end{matrix}\rs(\tau) &=& \sum_{n,m\in\Z}\lp -1\rp^{n+m+1}q_1^{\hlf\lp n+\hlf\rp^2}q_2^{\hlf\lp m+\hlf\rp^2}q_3^{\hlf\lp n+m+1\rp^2}.
\eea
In each of these expressions the exponents of each $q_i$ are either always integer or half-integer, indicating that in this contribution the corresponding tube has $|1\rangle$ or $|\epsilon\rangle$ states or their descendants propagating, or they are integer plus one-eighth, indicating that the $|\s\rangle$ states and its descendants are in play.  Thus, we are led to propose the following expressions for the untwisted partial traces (now leaving $\tau$ arguments implicit on theta functions),
\bea
Z_{1,1;1,1} &=& Z^{(2)}=H_2(\tau)\left\{\left|\theta\ls\begin{matrix}0 & 0 \\ 0 & 0\end{matrix}\rs\right|+\left|\theta\ls\begin{matrix}0 & 0 \\ 0 & \hlf\end{matrix}\rs\right|+\left|\theta\ls\begin{matrix}0 & 0 \\ \hlf & 0\end{matrix}\rs\right|+\left|\theta\ls\begin{matrix}0 & 0 \\ \hlf & \hlf\end{matrix}\rs\right|\right.\non\\
&& \qquad\left. +\left|\theta\ls\begin{matrix}\hlf & 0 \\ 0 & 0\end{matrix}\rs\right|+\left|\theta\ls\begin{matrix}\hlf & 0 \\ 0 & \hlf\end{matrix}\rs\right|+\left|\theta\ls\begin{matrix}0 & \hlf \\ 0 & 0\end{matrix}\rs\right|+\left|\theta\ls\begin{matrix}0 & \hlf \\ \hlf & 0\end{matrix}\rs\right|\right.\non\\
&& \qquad\left. +\left|\theta\ls\begin{matrix}\hlf & \hlf \\ 0 & 0\end{matrix}\rs\right|+\left|\theta\ls\begin{matrix}\hlf & \hlf \\ \hlf & \hlf\end{matrix}\rs\right|\right\},\\
Z_{1,1;1,-1} &=& H_2(\tau)\left\{\left|\theta\ls\begin{matrix}0 & 0 \\ 0 & 0\end{matrix}\rs\right|+\left|\theta\ls\begin{matrix}0 & 0 \\ 0 & \hlf\end{matrix}\rs\right|+\left|\theta\ls\begin{matrix}0 & 0 \\ \hlf & 0\end{matrix}\rs\right|+\left|\theta\ls\begin{matrix}0 & 0 \\ \hlf & \hlf\end{matrix}\rs\right|\right.\non\\
&& \qquad\left. +\left|\theta\ls\begin{matrix}\hlf & 0 \\ 0 & 0\end{matrix}\rs\right|+\left|\theta\ls\begin{matrix}\hlf & 0 \\ 0 & \hlf\end{matrix}\rs\right|-\left|\theta\ls\begin{matrix}0 & \hlf \\ 0 & 0\end{matrix}\rs\right|-\left|\theta\ls\begin{matrix}0 & \hlf \\ \hlf & 0\end{matrix}\rs\right|\right.\non\\
&& \qquad\left. -\left|\theta\ls\begin{matrix}\hlf & \hlf \\ 0 & 0\end{matrix}\rs\right|-\left|\theta\ls\begin{matrix}\hlf & \hlf \\ \hlf & \hlf\end{matrix}\rs\right|\right\},\\
Z_{1,1;-1,1} &=& H_2(\tau)\left\{\left|\theta\ls\begin{matrix}0 & 0 \\ 0 & 0\end{matrix}\rs\right|+\left|\theta\ls\begin{matrix}0 & 0 \\ 0 & \hlf\end{matrix}\rs\right|+\left|\theta\ls\begin{matrix}0 & 0 \\ \hlf & 0\end{matrix}\rs\right|+\left|\theta\ls\begin{matrix}0 & 0 \\ \hlf & \hlf\end{matrix}\rs\right|\right.\non\\
&& \qquad\left. -\left|\theta\ls\begin{matrix}\hlf & 0 \\ 0 & 0\end{matrix}\rs\right|-\left|\theta\ls\begin{matrix}\hlf & 0 \\ 0 & \hlf\end{matrix}\rs\right|+\left|\theta\ls\begin{matrix}0 & \hlf \\ 0 & 0\end{matrix}\rs\right|+\left|\theta\ls\begin{matrix}0 & \hlf \\ \hlf & 0\end{matrix}\rs\right|\right.\non\\
&& \qquad\left. -\left|\theta\ls\begin{matrix}\hlf & \hlf \\ 0 & 0\end{matrix}\rs\right|-\left|\theta\ls\begin{matrix}\hlf & \hlf \\ \hlf & \hlf\end{matrix}\rs\right|\right\},\\
Z_{1,1;-1,-1} &=& H_2(\tau)\left\{\left|\theta\ls\begin{matrix}0 & 0 \\ 0 & 0\end{matrix}\rs\right|+\left|\theta\ls\begin{matrix}0 & 0 \\ 0 & \hlf\end{matrix}\rs\right|+\left|\theta\ls\begin{matrix}0 & 0 \\ \hlf & 0\end{matrix}\rs\right|+\left|\theta\ls\begin{matrix}0 & 0 \\ \hlf & \hlf\end{matrix}\rs\right|\right.\non\\
&& \qquad\left. -\left|\theta\ls\begin{matrix}\hlf & 0 \\ 0 & 0\end{matrix}\rs\right|-\left|\theta\ls\begin{matrix}\hlf & 0 \\ 0 & \hlf\end{matrix}\rs\right|-\left|\theta\ls\begin{matrix}0 & \hlf \\ 0 & 0\end{matrix}\rs\right|-\left|\theta\ls\begin{matrix}0 & \hlf \\ \hlf & 0\end{matrix}\rs\right|\right.\non\\
&& \qquad\left. +\left|\theta\ls\begin{matrix}\hlf & \hlf \\ 0 & 0\end{matrix}\rs\right|+\left|\theta\ls\begin{matrix}\hlf & \hlf \\ \hlf & \hlf\end{matrix}\rs\right|\right\},
\eea
Adding together all four of these we have
\begin{multline}
Z_{1,1}=\frac{1}{4}\lp Z_{1,1;1,1}+Z_{1,1;1,-1}+Z_{1,1;-1,1}+Z_{1,1;-1,-1}\rp\\
=H_2(\tau)\left\{\left|\ls\begin{matrix}0 & 0 \\ 0 & 0\end{matrix}\rs\right|+\left|\theta\ls\begin{matrix}0 & 0 \\ 0 & \hlf\end{matrix}\rs\right|+\left|\theta\ls\begin{matrix}0 & 0 \\ \hlf & 0\end{matrix}\rs\right|+\left|\theta\ls\begin{matrix}0 & 0 \\ \hlf & \hlf\end{matrix}\rs\right|\right\},
\end{multline}
which is just the result of restricting to invariant states in each tube.

Now, from the expressions in section \ref{g2mapping}, we can identify the $\Sp(4,\Z)$ matrices associated to various Dehn twists,
\bea
\label{g2mcgpres1}
D_{a_1} &:& A=D=1,\qquad C=0,\qquad B=\lp\begin{smallmatrix}1 & 0 \\ 0 & 0\end{smallmatrix}\rp,\\
D_{a_2} &:& A=D=1,\qquad C=0,\qquad B=\lp\begin{smallmatrix}0 & 0 \\ 0 & 1\end{smallmatrix}\rp,\\
D_{b_1} &:& A=D=1,\qquad B=0,\qquad C=\lp\begin{smallmatrix}-1 & 0 \\ 0 & 0\end{smallmatrix}\rp,\\
D_{b_2} &:& A=D=1,\qquad B=0,\qquad C=\lp\begin{smallmatrix}0 & 0 \\ 0 & -1\end{smallmatrix}\rp,\\
\label{g2mcgpres2}
D_c &:& A=D=1,\qquad C=0,\qquad B=\lp\begin{smallmatrix}1 & -1 \\ -1 & 1\end{smallmatrix}\rp.
\eea
Thus the action on $|\theta[\begin{smallmatrix}\al_1 & \al_2 \\ \beta_1 & \beta_2\end{smallmatrix}]|$ is
\bea
D_{a_1}\cdot\left|\theta\ls\begin{matrix}\al_1 & \al_2 \\ \beta_1 & \beta_2\end{matrix}\rs\right| &=& \left|\theta\ls\begin{matrix}\al_1 & \al_2 \\ \beta_1-\al_1+\hlf & \beta_2\end{matrix}\rs\right|,\\
D_{a_2}\cdot\left|\theta\ls\begin{matrix}\al_1 & \al_2 \\ \beta_1 & \beta_2\end{matrix}\rs\right| &=& \left|\theta\ls\begin{matrix}\al_1 & \al_2 \\ \beta_1 & \beta_2-\al_2+\hlf\end{matrix}\rs\right|,\\
D_{b_1}\cdot\left|\theta\ls\begin{matrix}\al_1 & \al_2 \\ \beta_1 & \beta_2\end{matrix}\rs\right| &=& \left|1-\tau_{11}\right|^{1/2}\left|\theta\ls\begin{matrix}\al_1+\beta_1-\hlf & \al_2 \\ \beta_1 & \beta_2\end{matrix}\rs\right|,\\
D_{b_2}\cdot\left|\theta\ls\begin{matrix}\al_1 & \al_2 \\ \beta_1 & \beta_2\end{matrix}\rs\right| &=& \left|1-\tau_{22}\right|^{1/2}\left|\theta\ls\begin{matrix}\al_1 & \al_2+\beta_2-\hlf \\ \beta_1 & \beta_2\end{matrix}\rs\right|,\\
D_c\cdot\left|\theta\ls\begin{matrix}\al_1 & \al_2 \\ \beta_1 & \beta_2\end{matrix}\rs\right| &=& \left|\theta\ls\begin{matrix}\al_1 & \al_2 \\ \beta_1-\al_1+\al_2+\hlf & \beta_2+\al_1-\al_2+\hlf\end{matrix}\rs\right|.
\eea
We can now easily confirm that all the untwisted sector partial traces are invariant under $D_{a_1}$, $D_{a_2}$, and $D_c$, as we might expect.  Under $D_{b_1}$ and $D_{b_2}$, $Z^{(2)}$ is invariant but the others are not, and begin to generate twisted sector partial traces.  In the $(1,-1)$ twisted sector,
\bea
Z_{1,-1;1,-1} &=& D_{b_2}\cdot Z_{1,1;1,-1}=H_2(\tau)\left\{ -\left|\theta\ls\begin{matrix}0 & 0 \\ 0 & 0\end{matrix}\rs\right|+\left|\theta\ls\begin{matrix}0 & 0 \\ 0 & \hlf\end{matrix}\rs\right|-\left|\theta\ls\begin{matrix}0 & 0 \\ \hlf & 0\end{matrix}\rs\right|+\left|\theta\ls\begin{matrix}0 & 0 \\ \hlf & \hlf\end{matrix}\rs\right|\right.\non\\
&& \qquad\left. -\left|\theta\ls\begin{matrix}\hlf & 0 \\ 0 & 0\end{matrix}\rs\right|+\left|\theta\ls\begin{matrix}\hlf & 0 \\ 0 & \hlf\end{matrix}\rs\right|+\left|\theta\ls\begin{matrix}0 & \hlf \\ 0 & 0\end{matrix}\rs\right|+\left|\theta\ls\begin{matrix}0 & \hlf \\ \hlf & 0\end{matrix}\rs\right|\right.\non\\
&& \qquad\left. +\left|\theta\ls\begin{matrix}\hlf & \hlf \\ 0 & 0\end{matrix}\rs\right|-\left|\theta\ls\begin{matrix}\hlf & \hlf \\ \hlf & \hlf\end{matrix}\rs\right|\right\},\\
Z_{1,-1;-1,-1} &=& D_{b_2}\cdot Z_{1,1;-1,-1}=H_2(\tau)\left\{ -\left|\theta\ls\begin{matrix}0 & 0 \\ 0 & 0\end{matrix}\rs\right|+\left|\theta\ls\begin{matrix}0 & 0 \\ 0 & \hlf\end{matrix}\rs\right|-\left|\theta\ls\begin{matrix}0 & 0 \\ \hlf & 0\end{matrix}\rs\right|+\left|\theta\ls\begin{matrix}0 & 0 \\ \hlf & \hlf\end{matrix}\rs\right|\right.\non\\
&& \qquad\left. +\left|\theta\ls\begin{matrix}\hlf & 0 \\ 0 & 0\end{matrix}\rs\right|-\left|\theta\ls\begin{matrix}\hlf & 0 \\ 0 & \hlf\end{matrix}\rs\right|+\left|\theta\ls\begin{matrix}0 & \hlf \\ 0 & 0\end{matrix}\rs\right|+\left|\theta\ls\begin{matrix}0 & \hlf \\ \hlf & 0\end{matrix}\rs\right|\right.\non\\
&& \qquad\left. -\left|\theta\ls\begin{matrix}\hlf & \hlf \\ 0 & 0\end{matrix}\rs\right|+\left|\theta\ls\begin{matrix}\hlf & \hlf \\ \hlf & \hlf\end{matrix}\rs\right|\right\},
\eea
\bea
Z_{1,-1;1,1} &=& D_{a_2}\cdot Z_{1,-1;1,-1}=H_2(\tau)\left\{\left|\theta\ls\begin{matrix}0 & 0 \\ 0 & 0\end{matrix}\rs\right|-\left|\theta\ls\begin{matrix}0 & 0 \\ 0 & \hlf\end{matrix}\rs\right|+\left|\theta\ls\begin{matrix}0 & 0 \\ \hlf & 0\end{matrix}\rs\right|-\left|\theta\ls\begin{matrix}0 & 0 \\ \hlf & \hlf\end{matrix}\rs\right|\right.\non\\
&& \qquad\left. +\left|\theta\ls\begin{matrix}\hlf & 0 \\ 0 & 0\end{matrix}\rs\right|-\left|\theta\ls\begin{matrix}\hlf & 0 \\ 0 & \hlf\end{matrix}\rs\right|+\left|\theta\ls\begin{matrix}0 & \hlf \\ 0 & 0\end{matrix}\rs\right|+\left|\theta\ls\begin{matrix}0 & \hlf \\ \hlf & 0\end{matrix}\rs\right|\right.\non\\
&& \qquad\left. +\left|\theta\ls\begin{matrix}\hlf & \hlf \\ 0 & 0\end{matrix}\rs\right|-\left|\theta\ls\begin{matrix}\hlf & \hlf \\ \hlf & \hlf\end{matrix}\rs\right|\right\},\\
Z_{1,-1;-1,1} &=& D_{a_2}\cdot Z_{1,-1;-1,-1}=H_2(\tau)\left\{\left|\theta\ls\begin{matrix}0 & 0 \\ 0 & 0\end{matrix}\rs\right|-\left|\theta\ls\begin{matrix}0 & 0 \\ 0 & \hlf\end{matrix}\rs\right|+\left|\theta\ls\begin{matrix}0 & 0 \\ \hlf & 0\end{matrix}\rs\right|-\left|\theta\ls\begin{matrix}0 & 0 \\ \hlf & \hlf\end{matrix}\rs\right|\right.\non\\
&& \qquad\left. -\left|\theta\ls\begin{matrix}\hlf & 0 \\ 0 & 0\end{matrix}\rs\right|+\left|\theta\ls\begin{matrix}\hlf & 0 \\ 0 & \hlf\end{matrix}\rs\right|+\left|\theta\ls\begin{matrix}0 & \hlf \\ 0 & 0\end{matrix}\rs\right|+\left|\theta\ls\begin{matrix}0 & \hlf \\ \hlf & 0\end{matrix}\rs\right|\right.\non\\
&& \qquad\left. -\left|\theta\ls\begin{matrix}\hlf & \hlf \\ 0 & 0\end{matrix}\rs\right|+\left|\theta\ls\begin{matrix}\hlf & \hlf \\ \hlf & \hlf\end{matrix}\rs\right|\right\}.
\eea
Adding up, we have
\be
Z_{1,-1}=H_2(\tau)\left\{\left|\theta\ls\begin{matrix}0 & \hlf \\ 0 & 0\end{matrix}\rs\right|+\left|\theta\ls\begin{matrix}0 & \hlf \\ \hlf & 0\end{matrix}\rs\right|\right\}.
\ee

Proceeding similarly for the other sectors, we find
\be
Z_{-1,1}=H_2(\tau)\left\{\left|\theta\ls\begin{matrix}\hlf & 0 \\ 0 & 0\end{matrix}\rs\right|+\left|\theta\ls\begin{matrix}\hlf & 0 \\ 0 & \hlf\end{matrix}\rs\right|\right\},
\ee
and
\be
Z_{-1,-1}=H_2(\tau)\left\{\left|\theta\ls\begin{matrix}\hlf & \hlf \\ 0 & 0\end{matrix}\rs\right|+\left|\theta\ls\begin{matrix}\hlf & \hlf \\ \hlf & \hlf\end{matrix}\rs\right|\right\}.
\ee

As expected, our final result returns the original theory, as in the genus one calculation:
\be
Z_{1,1}+Z_{1,-1}+Z_{-1,1}+Z_{-1,-1}=Z^{(2)}.
\ee

%\newpage
%%%%%%%%%%%%%%%%%%%%%%%%%%%%%%%%%%%%%%%%%%%%%%%%%%%%%%%%%%%%

%\bibliographystyle{ieeetr}
%\bibliography{HigherGenusPaper}

%\providecommand{\href}[2]{#2}\begingroup\raggedright\begin{thebibliography}{10}

%\end{thebibliography}\endgroup

\end{document}